\documentclass[%
preprint,
 amsmath,amssymb,
 aps,
]{revtex4-1}

\usepackage{graphicx}
\usepackage{dcolumn}
\usepackage{bm}
\usepackage{ifpdf}
\usepackage{epstopdf}
\usepackage{slashed}
\usepackage{amsfonts}
\usepackage{mathrsfs}
\usepackage{amssymb}
\usepackage{multirow}
\usepackage{CJK}
\usepackage{xcolor}
\usepackage{lineno}
\usepackage{soul}
\usepackage{booktabs}
\soulregister\cite7
\soulregister\citep7
\soulregister\citet7
\soulregister\ref7
\soulregister\pageref7

\begin{document}

\title{A tensor decomposition model for evaluating isotopic yield in neutron-induced fission}% Force line

\author{Qufei Song}%
\author{Long Zhu}
\author{Jun Su}
\email{Corresponding author: sujun3@mail.sysu.edu.cn}

\affiliation{%
 Sino-French Institute of Nuclear Engineering and Technology, Sun Yat-sen University, Zhuhai 519082, China
}%

\author{Hui Guo}
\affiliation{%
 School of Nuclear Science and Engineering, Shanghai Jiao Tong University, Shanghai 200240, China}

\date{\today}% It is always \today, today,
             %  but any date may be explicitly specified
\begin{abstract}
	\begin{description}
		\item[Background] Due to the complex multidimensional dependence, the prediction and evaluation of independent fission yield distributions have always been a challenge.
		
		\item[Purpose] Considering the complex multidimensional dependence and high missing rate of independent yield data, this work applies the tensor decomposition algorithm to the prediction of independent fission yields.
		
		\item[Methods] After constructing yield tensors with three dimensions for 851 fission products and filling the tensors with the independent yield data from the ENDF/B-VIII.0 database, the tensor decomposition algorithm is applied to predict the independent isotopic yield in fission, which resulting in the Fission Yield Tensor Decomposition (FYTD) model.
		
		\item[Results] The fission yields of $^{235}$U and $^{239}$Pu are set as missing values and then predicted.
The predictions for $^{235}$U fissions by the FYTD model agree with the ENDF/B-VIII.0 data and are better than those by the Talys and BNN+Talys models.
Furthermore, we predict not only the mass distribution but also the isotopic yields in the fissions.
For fast neutron-induced fission of $^{239}$Pu, 98\% predictions of the isotopic yields by the FYTD model agree with the ENDF/B-VIII.0 data within 1 order of magnitude.
The fission yields of $^{238}$Np, $^{243}$Am, and $^{236}$Np that do not exist in the ENDF/B-VIII.0 database are predicted and compared with those in the JEFF-3.3 database, as well as the experimental data.
Good agreement demonstrates the predictive ability of the FYTD model for the target nucleus dependence.
The scalability of the yield tensor decomposition model over the incident neutron energy degrees of freedom is examined.
After adding a set of 2 MeV neutron-induced $^{239}$Pu fission yield data into the yield tensor, the 2 MeV neutron-induced fission yields of $^{235}$U and $^{239}$Pu are predicted.
The comparison with the experimental data shows that the predictions are similar to those by the GEF model in the peak area but more accurate in the valley area.
Finally, the yields of the ratio of isomeric states and neutron excess of the products as a function of product charge number are also studied.
		
		\item[Conclusions] The FYTD model can capture the multi-dimensional dependence of the fission yield data and make reasonable predictions. The FYTD model has scalability in the energy dimension and can predict the yields of the ratio of isomeric states.
	\end{description}
\end{abstract}

\maketitle
%\pagewiselinenumbers% 按页重新编号

\section{\label{int}Introduction}

The splitting of a heavy nucleus into two or more intermediate-mass nuclei is called fission.  Although it has been more than 80 years since its discovery \cite{hahn1939nachweis, meitner1939disintegration}, the research on the fission is still a hot topic and challenge.
On the one hand, the nuclear fission is an extremely complex process, which is the movement of the quantum multi-body system composed of all nucleons in the nucleus in the multidimensional space.
The exploration of its mechanism is very helpful to the development of nuclear physics fields such as nuclear structure, nuclear reaction and super-heavy nuclear research \cite{schunck2016microscopic, bender_future_2020, hamilton_search_2013, pei_fission_2009}.The nuclear fission also attracts attentions in astrophysics and particle physics because it plays an important role in the formation of elements in the rapid neutron capture process (r-process) of nucleosynthesis and the production of reactor neutrinos \cite{eichler_role_2015, mueller_improved_2011}.
On the other hand, the nuclear fission is also significant application fields.
The huge energy released in the fission process makes fission play an important role in both energy and military fields.
Neutrons and various radioisotopes produced by fission are used in various fields such as biology, chemistry and medicine.
Therefore, in order to make scientific use of fission, the research on the nuclear fission is widely concerned in the field of nuclear engineering and technology \cite{bernstein_our_2019}.

The fission product yield (FPY) is an important observable in the nuclear fission.
Generally speaking, the nuclear fission is followed by the $\beta$ decay of the unstable fragments.
The FPY is divided into the independent and cumulative cases, which are distinguished by counting the products before or after the $\beta$ decay.
The independent fission product yield (IFPY) can reflect the information of fission process from the macro and micro perspectives, provide important observations for the research and modeling of fission process \cite{ramos_insight_2019}.
However, experimental measurements of the IFPY are difficult and hence the available data is generally incomplete and have large uncertainties \cite{denschlag_independent_1986}.
In major nuclear data libraries, such as ENDF/B \cite{brown_endfb-viii0_2018}, JEFF \cite{kellett_jeff-31-311_2009}, and JENDL \cite{shibata_jendl-40_2011}, complete evaluations of IFPY are not available for some certain actinides and only available for three neutron incident energy points (0.0253 eV, 0.5 MeV and 14 MeV).
Therefore, theoretical predictions of the IFPY are still necessary.

Due to the complexity of the quantum many body problem and the nuclear force problem, the deep understanding and simulation of the fission process is still one of the most challenging tasks in nuclear physics \cite{schunck2016microscopic, bender_future_2020}.
Nowadays, the microscopic nuclear fission models, such as the time-dependent Hartree-Fock-Bogoliubov method \cite{bulgac_induced_2016} and the time-dependent generator coordinate method \cite{regnier_fission_2016, younes_microscopic_2019}, have made important progress.
The fission process can be regarded as a movement of Brownian particles walking on the multi-dimensional potential energy surface.
Various macroscopic-microscopic models based on the multi-dimensional potential energy surface are also widely used in the calculation of the fission yield \cite{randrup_brownian_2011, randrup_fission-fragment_2011, pomorski_mass_2017, liu_study_2019}.
For practical applications requiring higher precision, phenomenological methods are more widely used. For example, the multi-Gaussian semi-empirical formula \cite{ramos_insight_2019, fang_theoretical_2021}, the Brosa model \cite{brosa_nuclear_1990} and the GEF (GEneral description of Fission observables) model \cite{schmidt_general_2016} have all achieved considerable success in evaluating fission yield data.
The prediction of the GEF model is also referenced in the JEFF database to further improve the fission yield data in the database.

The traditional phenomenological models mainly rely on the least-squares adjustment of various parameters.
They can describe the existing data in some regions well.
But as fission modes evolve, the prediction capability of this kind of models may be insufficient when the available experimental data are very sparse \cite{schmidt2018review, wang_bayesian_2019}. Recently, thanks to the powerful ability to learn from existing data and make predictions, the machine learning algorithms have been used in various studies in the nuclear physics community.
For example, the Bayesian neural network (BNN) algorithm has achieved certain success in prediction of nuclear mass \cite{niu_nuclear_2018, utama_nuclear_2016-1}, nuclear charge radius \cite{utama_nuclear_2016}, spallation reaction product cross section \cite{ma_isotopic_2020, song2022target}, neutron and proton drip lines \cite{neufcourt_neutron_2019, neufcourt_beyond_2020}, etc. Likewise, various machine learning algorithms have been applied to the study and prediction of fission yields.
Lovell et al. used Mixture Density Networks to learn parameters of Gaussian functions to predict fission product yields \cite{lovell_constraining_2019}.
Wang et al. predicted the mass distribution of fission yield by combining BNN and talys models \cite{wang_bayesian_2019}. Qiao et al. used BNN model to predict the charge distribution of $^{239}$Pu fission yield \cite{qiao_bayesian_2021}.

The tensor decomposition algorithm is a standard technique to capture the multi-dimensional structural dependence.
%The basic idea of this algorithm is to use mathematical methods to decompose the tensor into multiple factor matrices, and then update the factor matrix according to the existing data in the tensor.
%After reconstituting the factor matrix into a tensor, the predicted values of missing elements in the tensor can be obtained.
Compared with the traditional interpolation and fitting methods, the tensor decomposition algorithm has a strong ability to extract the information hidden in the original data, making it possible to impute the sparse tensor, which makes it widely used in image processing, data mining and other fields \cite{liu_tensor_2013, chen_bayesian_2019, chen_general_2015}.
%The tensor decomposition is not yet widely used in nuclear physics.
Considering the complex multidimensional dependence and high missing rate of independent yield data, this work applied the tensor decomposition algorithm to the prediction of independent fission yields, and established the Fission Yield Tensor Decomposition (FYTD) model.
The paper is organized as follows. In Sec. \ref{method}, the establishment of FYTD model is described. In Sec. \ref{results}, FYTD model will be applied to multiple yield prediction for verification. Finally, Sec. \ref{summary} presents conclusions and perspectives for future studies.

\section{\label{method}Theoretical framework}

%In the FYTD model, the fission yield data were firstly tensorized,  constructing yield
%tensors with three dimensions for 851 fission products
%Second, the constructed yield tensor is filled with the independent yield data from the ENDF/B-VIII.0 database to obtain the missing tensor.
%The yield tensor is then decomposed into three factor matrices, and the factor matrices are iteratively updated using the Bayesian Gaussian CANDECOMP/PARAFAC (BGCP) tensor decomposition model algorithm \cite{chen_bayesian_2019}.
%After reconstruction of three factor matrices, the results are normalized to obtain the final prediction result.

\subsection{Tensorization of fission product yield}
The neutron-induced fission is briefly introduced by taking a typical $^{235}$U(n,f) reaction as an example:
\begin{equation}
	 ^{235}_{92}\text {U} + \text {n} \longrightarrow ^{236}_{92}\text {U}^{*} \longrightarrow ^{92}_{36}\text {Kr} + ^{141}_{56}\text {Ba} + \text {3n}.
\end{equation}
A specific target nucleus, such as $^{235}$U, can be represented by its number of protons $Z_{t}$ and number of neutrons $N_{t}$. The target nucleus forms an excited composite nucleus after the neutron incident with energy $E_{n}$. After fission of composite nucleus, two primary fission products, one light and one heavy, are produced and several prompt neutrons are released.
Thus, the neutron-induced fission yield of a given isotope with number of protons $Z_{p}$ and number of neutrons $N_{p}$ depends on the neutron incident energy and target nuclei. In addition, a considerable part of the fission products are not in the ground state but in the isomeric state. In order to distinguish these products, the Fission Product State (FPS) needs to be considered.

Therefore, when tensorizing the neutron-induced fission product yield data, six dimensions can be considered, including $E_{n}$, $Z_{t}$, $N_{t}$, $Z_{p}$, $N_{p}$ and FPS. Among six seven dimensions, the dimensions related to the target nucleus and product are discrete, and only the neutron incident energy $E_{n}$ is continuous. Therefore, $E_{n}$ is discretized into thermal neutron (0.0253 eV), fast neutron (0.5 MeV) and high-energy neutron (14 MeV).

Nowadays, due to the difficulty of IFPY experimental measurement and the lack of existing data, this yield tensor will have a very high degree of missingness. Imputation of such sparse tensors has always been a challenge. Similar scenes also appear in image processing in the computer field. A color image can be thought of as a multidimensional tensor containing information about the pixel location dimension, as well as the pixel color dimension. In order to deal with color images with high missing rate and high signal-to-noise ratio, tensor decomposition algorithms are widely used \cite{liu_tensor_2013, chen_bayesian_2019, chen_general_2015}. Taking the image in the upper left corner of Fig. \ref{framework} as an example, 70$\%$ of the pixels in this image are missing. We tensorized the image and applied the BGCP algorithm to capture the multidimensional information of the remaining pixels. Finally, as shown in the image in the upper right corner of Fig. \ref{framework}, tensor decomposition can effectively impute the missing pixels.

However, it is not feasible to directly build a 6D tensor and impute it. Even based on the ENDF/B-VIII.0 database with the most abundant IFPY data, the missing rate of this 6D tensor can still be as high as 98 $\%$. Excessive tensor construction range will not only bring about a high missing rate, but also incorporate some non-existent extreme neutron-rich nuclei, proton-rich nuclei, and isomeric states into the tensor, which will make some elements of the tensor physically meaningless.
Taking into account this problem, this work proposes a method for constructing 3D tensors.
The 3D yield tensors with dimensions $Z_{t}$, $N_{t}-Z_{t}$ and $E_{n}$ were constructed separately for 851 products with relatively rich existing data in ENDF/B-VIII.0 database. The range of the tensor is chosen as $Z_{t}$ = 90-96, $N_{t}-Z_{t}$ = 47-54 and $E_{n}$ = 0.0253 eV, 0.5 MeV, 14 MeV. By filling these 3D tensors with the ENDF/B-VIII.0 data of 45 fission systems for 24 target nuclei of $^{227, 229, 232}$Th, $^{231}$Pa, $^{232, 233, 234, 236, 237, 238}$U, $^{237, 238}$Np,  $^{238, 239, 240, 241, 242}$Pu, $^{241, 243}$Am and $^{242, 243, 245, 246}$Cm, the missing rate can be reduced to 73$\%$, and the tensor will not contain any non-existing target nuclei, product nuclei and isomeric states.

Now, let $Y_{i j k}$ represent the exciting yield in ENDF/B-VIII.0 database, where $i,j,k$ represents $Z_{t}$, $N_{t}-Z_{t}$ and $E_{n}$ respectively. For example, $Y_{3,5,1}$ is entry with $i = 3 (Z_{t} = 92)$, $j = 5 (N_{t}-Z_{t} = 51)$ and $k = 1 (E_{n} = 0.0253 eV)$, which represents yield in 0.0253 eV neutron induced fission of $^{235}$U. The magnitude of the independent yield is very small, even up to 10$^{-18}$. Therefore, it is necessary to fill the tensor with the logarithm of the yield, in case this data is too small for numerical calculation. This approach works well for products with big variation, allowing the algorithm to capture magnitude changes in yield well.
But for products with small yield variation, the difference is even smaller after logarithmization, and it is difficult for the algorithm to capture their difference. Therefore, to solve this problem, we analyzed the degree of dispersion of 851 product yield data under different fission systems. In this work the degree of dispersion is defined as the standard deviation of the logarithm values of the existing yield data:

\begin{equation}
	\sigma_{\text {yield }}=\sqrt{\sum_{i=1}^{I} \sum_{j=1}^{J} \sum_{k=1}^{K} b_{i j k}\left[\frac{\ln \left(Y_{i j k}\right)-\overline{\ln (Y)}}{\sum_{i=1}^{I} \sum_{j=1}^{J} \sum_{k=1}^{K} b_{i j k}}\right]},
\end{equation}
where $b_{i j k}$ is 1 for the exciting yield data and 0 for the missing yield data. If $\sigma_{\text {yield }} > 1$, its magnitude varies greatly, and the yield is logarithmized during filling; otherwise, the yield remains linear.

At this point, 851 yield tensors containing ENDF/B-VIII.0 data and missing elements were obtained. Let $\mathcal{Y}$ represent these missing tensor and $\hat{\mathcal{Y}} \in \mathbb{R}^{I \times J \times K}$ the physical reality of the yield data. In fact, the physical reality value $\hat{Y}_{i j k}$ cannot be known, only some exciting observations with uncertainties $Y_{i j k}$ are available. In this work, BGCP tensor decomposition model algorithm \cite{chen_bayesian_2019} will be applied to estimate the physical reality $\hat{\mathcal{Y}}$ according to the existing observation $\mathcal{Y}$.

\subsection{Bayesian Gaussian CANDECOMP/PARAFAC tensor decomposition}
A detailed introduction to the Bayesian Gaussian CANDECOMP/PARAFAC (BGCP) algorithm can be found in Ref. \cite{chen_bayesian_2019}. Here is a brief introduction to the application of BGCP in this work. It is assumed that the uncertainty of each exciting yield data $Y_{i j k}$ follows an independent Gaussian distribution,
\begin{equation}
Y_{i j k} \sim \mathcal{N}\left(\hat{Y}_{i j k}, \tau_{\epsilon}^{-1}\right),
\label{1}
\end{equation}
where $\tau_{\epsilon}$ is the precision. In real-world applications the expectation of yield $\hat{Y}_{i j k}$ is unknown and replaced with the estimated yield, which is the entry of the estimated tensor $\hat{\mathcal{Y}}$. The CP decomposition is applied to calculate the estimation $\hat{\mathcal{Y}}$ :
\begin{equation}
\hat{\mathcal{Y}}=\sum_{n=1}^{r} \boldsymbol{z}^{(n)} \circ \boldsymbol{d}^{(n)} \circ \boldsymbol{e}^{(n)},
\label{2}
\end{equation}
where $\boldsymbol{z}^{(n)} \in \mathbb{R}^{I}, \boldsymbol{d}^{(n)} \in \mathbb{R}^{J}$, and $\boldsymbol{e}^{(n)} \in \mathbb{R}^{K}$ are respectively the $n$-th column vector of the factor matrices $\boldsymbol{Z} \in \mathbb{R}^{I \times r}$, $\boldsymbol{D} \in \mathbb{R}^{J \times r}$, and $\boldsymbol{E} \in \mathbb{R}^{K \times r}$. The symbol $\circ$ represents the outer product.

The prior distribution of the row vectors of the factor matrix $\boldsymbol{Z}$ is the multivariate Gaussian
\begin{equation}
z_{i} \sim \mathcal{N}\left[\boldsymbol{\mu}_{i}^{(z)},\left(\boldsymbol{\Lambda}_{i}^{(z)}\right)^{-1}\right]
\label{3}
\end{equation}
where the hyper-parameter $\boldsymbol{\mu}^{(z)} \in \mathbb{R}^{r}$ expresses the expectation, and $\boldsymbol{\Lambda}^{(z)} \in \mathbb{R}^{r \times r}$ indicates the width of the distribution. The likelihood function can be written as
\begin{equation}
\mathcal{L}\left(Y_{i j k} \mid z_{i}, \boldsymbol{d}_{j}, \boldsymbol{e}_{k}, \tau_{\epsilon}\right) \propto \exp \left\{-\frac{\tau_{\epsilon}}{2}\left[Y_{i j k}-\left(z_{i}\right)^{T}\left(\boldsymbol{d}_{j} \circledast \boldsymbol{e}_{k}\right)\right]^{2}\right\}
\label{4}
\end{equation}
where $\circledast$ is the Hadamard product.

Then, taking $z_ {i}$ as an example, according to Bayesian theorem, the posterior distribution of $z_{i}$ after observing $Y_{i j k}$ is:
\begin{equation}
\begin{aligned}
	\mathcal{L}\left(z_{i} \mid Y_{i j k}, d_{j}, e_{k}, \tau_{\varepsilon}\right) & \propto \mathcal{L}\left(Y_{i j k} \mid z_{i}, d_{j}, e_{k}, \tau_{\varepsilon}\right) \times \operatorname{Pr}\left(z_{i}\right) \\
	&\propto \exp \left\{-\frac{\tau_{\varepsilon}}{2}\left[Y_{i j k}-\left(z_{i}\right)^{T}\left(d_{j} \circledast e_{k}\right)\right]^{2}\right\} \times \mathcal{N}\left[\mu_{i}^{(z)},\left(\Lambda_{i}^{(z)}\right)^{-1}\right].
\end{aligned}
\end{equation}

Then the posterior values of the hyper-parameters $\boldsymbol{\mu}^{(z)}$ and $\boldsymbol{\Lambda}^{(z)}$ are given as
\begin{equation}
\begin{gathered}
	\widehat{\Lambda}_{i}^{(z)}=\Lambda_{i}^{(z)}+\Delta \Lambda_{i}^{(z)}, \\
	\Delta \Lambda_{i}^{(z)}=\tau_{\varepsilon}\left(d_{j} \otimes e_{k}\right)\left(d_{j}^{\otimes} e_{k}\right)^{T} . \\
	\hat{\mu}_{i}^{(z)}=\mu_{i}^{(z)}+\Delta \mu_{i}^{(z)}, \\
	\Delta \mu_{i}^{(z)}=\left(\widehat{\Lambda}_{i}^{(z)}\right)^{-1}\left(d_{j} \circledast e_{k}\right) \tau_{\varepsilon}\left[Y_{i j k}-\left(d_{j} \circledast e_{k}\right)^{T} \mu_{i}^{(z)}\right] .
\end{gathered}
\label{5}
\end{equation}
The contribution of the exciting yield data to the hyper-parameter is equivalent, and likelihood function of all exciting yield data is
\begin{equation}
\begin{aligned}
	\mathcal{L}\left(\mathcal{Y} \mid \boldsymbol{Z}, \boldsymbol{D}, \boldsymbol{E}, \tau_{\epsilon}\right) \propto &  \prod_{i=1}^{I} \prod_{j=1}^{J} \prod_{k=1}^{K}\left(\tau_{\epsilon}\right)^{1 / 2} \\
	& \times \exp \left[-\frac{\tau_{\epsilon}}{2} b_{i j k}\left(Y_{i j k}-\hat{Y}_{i j k}\right)^{2}\right]
\end{aligned}
\label{6}
\end{equation}
where $b_{i j k}$ is 1 for the exciting yield data and 0 for the missing yield data. Placing a conjugate $\Gamma$ prior to the precision $\tau_{\epsilon}$,
\begin{equation}
\tau_{\epsilon} \sim \Gamma\left(a_{0}, b_{0}\right)
\label{7}
\end{equation}
The posterior values of the hyper-parameters $a_{0}$ and $b_{0}$ are given as
\begin{equation}
\begin{aligned}
	&\hat{a}_{0}=\frac{1}{2} \sum_{i=1}^{I} \sum_{j=1}^{J} \sum_{k=1}^{K} b_{i j k}+a_{0} \\
	&\hat{b}_{0}=\frac{1}{2} \sum_{i=1}^{I} \sum_{j=1}^{J} \sum_{k=1}^{K}\left(Y_{i j k}-\hat{Y}_{i j k}\right)^{2}+b_{0}
\end{aligned}
\label{8}
\end{equation}
Based on Eq. (\ref{8}), each exciting yield data contributes to the increase of $\frac{1}{2}$ in $\hat{a}_{0}$, and $\frac{1}{2}\left(Y_{i j k}-\hat{Y}_{i j k}\right)^{2}$ in $\hat{b}_{0}$. Cases for the subscripts $j$ and $k$ are similar.

\begin{figure*}
	\centering
	\includegraphics[width=17.2cm,angle=0]{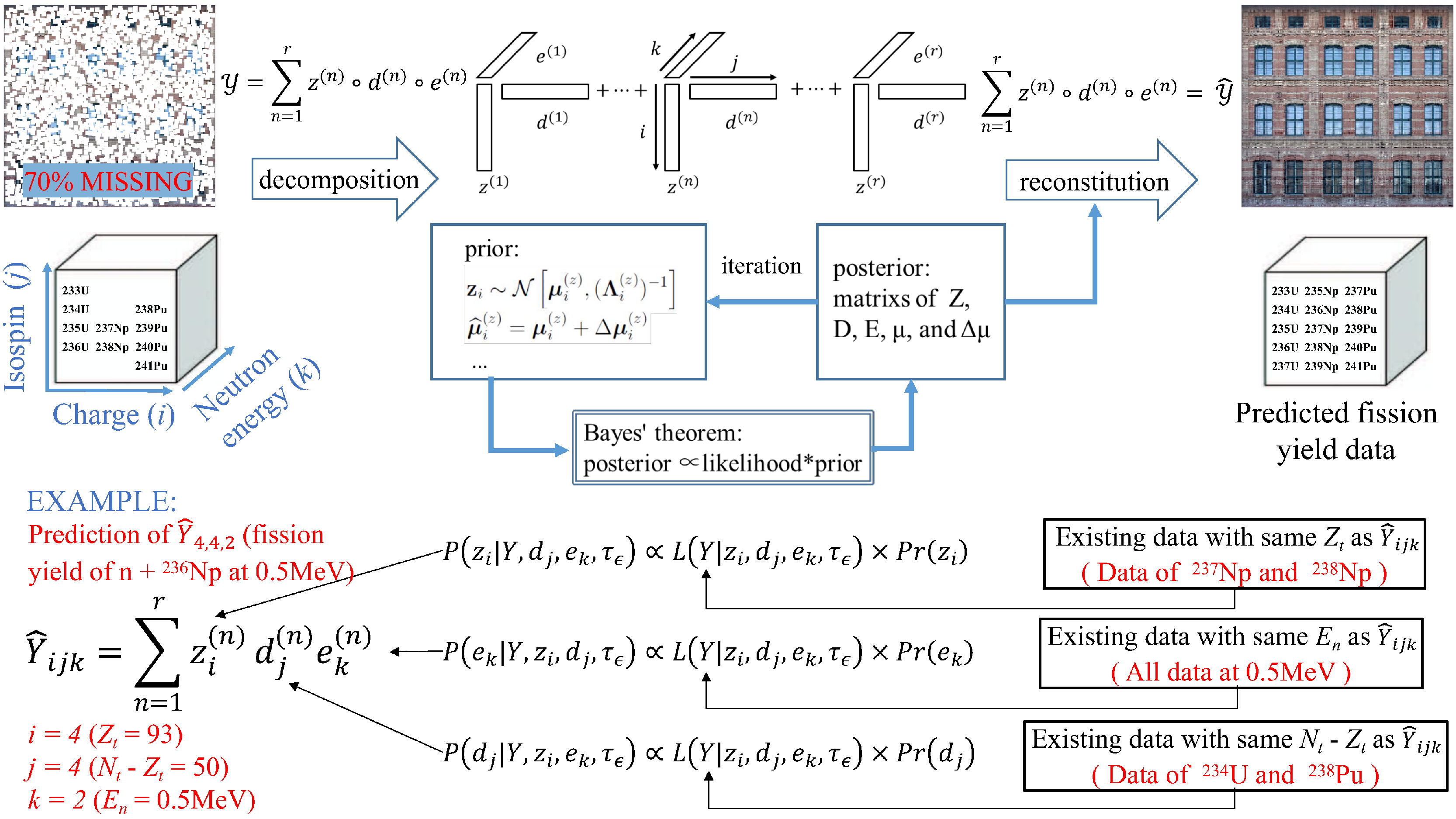}
	\caption{\label{framework}
		(Color online) Model framework presentation and example.
	}
\end{figure*}
In Fig. \ref{framework}, the above method is illustrated and a example is given. In brief, for a specific fission product, we denote its yields in different fission systems by $\hat{\mathcal{Y}}$. According to the $\mathrm{CP}$ decomposition, the tensor $\hat{\mathcal{Y}}$ is expressed as the outer product of the factor matrices $\boldsymbol{Z}, \boldsymbol{D}$, and $\boldsymbol{E}$. The prior distributions of the factor matrices are assumed to be multivariate Gaussians. With the exciting observed yield data, the posterior values of the factor matrices and their distributions can be calculated using Bayesian inference and iteration. Taking the prediction of $^{236}$Np fission yield induced by 0.5 MeV neutrons as an example, the information captured by the algorithm comes from the existing data on the same dimension, the same $Z_{t}$ data ($^{237}$Np and $^{238}$Np data), the same $N_{t}-Z_{t}$ data ($^{234}$U and $^{238}$Pu yield data) and all yield data under 0.5 MeV. Finally, the predicted fission yield is reconstituted with the factor matrices $\boldsymbol{Z}$, $\boldsymbol{D}$, and $\boldsymbol{E}$.

Without considering ternary fission, the products produced by each fission should be two kinds, so the sum of all fission product yields should also be 2. However, after tensor reconstruction, the sum of the yields obtained is not strictly 2 due to the precision of numerical calculation, and the yield of completion is always slightly smaller, such as 1.96, 1.98, etc. Therefore, it is necessary to perform a certain physical correction on the value. In this paper, the mass of the fission system minus the number of prompt neutrons is divided by two, and this value is used as the standard to judge whether the fission products belong to light nuclei or heavy nuclei, and the yields of light nuclei and heavy nuclei are normalized to 1, respectively. After normalization, the predicted yield data is finally obtained.

\section{\label{results}Results and discussions}

In order to quantitatively evaluate the prediction of FYTD model, the Root Mean Square Error (RMSE) and $\chi_{N}^{2}$ are used to measure the deviation between prediction and the ENDF/B-VIII.0 data. For the convenience of presentation, ENDF/B-VIII.0 is abbreviated as ENDF/B in the following text and figures.
For a fission system, the RMSE is calculated by evaluating the deviation of the predicted results of its 851 products from the ENDF/B:
\begin{equation}
RMSE=\sqrt{\frac{1}{N} \sum_{p=1}^{N}\left[log(\widehat{Y}^{p})-log(Y^{p})\right]^{2}},
\end{equation}
where N = 851, $\widehat{Y}^{p}$ represents the FYTD model prediction of the yield of the p-th product and $Y^{p}$ represents the corresponding ENDF/B data.

Considering most current fission yield prediction works and experimental measurements mainly focus on the mass distribution of the product, in order to compare with other models and data, as defined in Ref. \cite{wang_bayesian_2019}, use $\chi_{N}^{2}$ to measure the deviation of the predicted value of the mass distribution from the ENDF/B data:
\begin{equation}
\chi_{N}^{2}=\frac{1}{N} \sum_{p=1}^{N}\left(\widehat{Y}^{p}-Y^{p}\right)^{2}
\end{equation}
Here N = 107, which means the range of statistics is $\mathrm{A}=66-172$ for a total of 107 mass points.

The evaluation of RMSE and $\chi_{N}^{2}$ have different emphases. In calculation of RMSE, the magnitude difference between the predicted value of each product and the ENDF/B data is considered, which can globally evaluate the accuracy of magnitude prediction. Therefore it does not ignore the contribution of some products with small yield values. In contrast, $\chi_{N}^{2}$ focuses more on evaluating the accuracy of peak area predictions.

\subsection{ENDF/B-VIII.0 data prediction}

In this part, the ENDF/B yield data for each fission system are sequentially removed and predicted by FYTD model to systematically analyze the learning and prediction ability of FYTD model. Table \ref{RMSE} shows the RMSE of the FYTD model when predicting each fission system. In general, it can be seen that after removing the data to be predicted, the more remaining data in the learning set with the same dimension as the data to be predicted, the smaller the RMSE. For example, the RMSE of the prediction data for U, Pu, and Cm are mostly small, because even if the learning data of a fission system is removed, there are still many existing data in the same dimension that can capture information during prediction. For $^{229}$Th and $^{232}$Th, after removing their data from the learning set, there are few existing data that can be referenced when predicting them, resulting in a high RMSE. It can be found that the largest RMSE occurs at $^{227}$Th and $^{232}$U. This is because after the data of $^{227}$Th was removed, the data of j = 1 ($N_{t}-Z_{t}$ = 47) does not exist in the learning set at all. Therefore, when making predictions, it is completely impossible to capture the information of this dimension, resulting in unreliable predictions and extremely large RMSE. We believe that such predictions lacking dimensional information is unreliable and should be avoided when predicting using FYTD model. And in all following predictions, the situation will not occur, all calculations will avoid this problem.

\begin{table}[!htbp]
	\centering
	\caption{\label{RMSE} ENDF/B-VIII.0 data contained in the learning set of this work and prediction RMSE}
	\begin{tabular}{cccc}
		\hline\hline
		RMSE  & 0.0253 eV & 0.5 MeV & 14 MeV   \\
		\hline
		$^{227}$Th & 2.034 &       &        \\
		$^{229}$Th & 1.095 &       &        \\
		$^{232}$Th &       & 0.942 & 0.866  \\
		$^{231}$Pa &       & 1.005 &        \\
		$^{232}$U  & 1.874 &       &        \\
		$^{233}$U  & 0.598 &	0.473 &	0.411  \\
		$^{234}$U  &       & 0.659 &	0.244  \\
		$^{235}$U  & 0.622 &	0.552 &	0.260  \\
		$^{236}$U  &       & 0.391 &	0.271  \\
		$^{237}$U  &       & 0.419 &        \\
		$^{238}$U  &       & 0.712 &	0.683  \\
		$^{237}$Np & 0.466 &	0.612 &	0.622  \\
		$^{238}$Np &       & 0.607 &        \\
		$^{238}$Pu &       & 0.768 &        \\
		$^{239}$Pu & 0.429 &	0.281 &	0.390  \\
		$^{240}$Pu & 0.284 &	0.326 &	0.554  \\
		$^{241}$Pu & 0.652 &	0.424 &        \\
		$^{242}$Pu & 0.451 &	0.482 &	1.170  \\
		$^{241}$Am & 0.560 &	0.495 &	0.498  \\
		$^{243}$Am &       & 0.638 &        \\
		$^{242}$Cm &       & 0.839 &        \\
		$^{243}$Cm & 0.683 & 0.521 &        \\
		$^{244}$Cm &       & 0.443 &        \\
		$^{245}$Cm & 0.779 &       &        \\
		$^{246}$Cm &       & 0.637 &        \\
		\hline\hline
	\end{tabular}
\end{table}

\subsection{$^{235}$U and $^{239}$Pu fission yield prediction}

\begin{figure}
	\centering
	\includegraphics[width=8.6cm,angle=0]{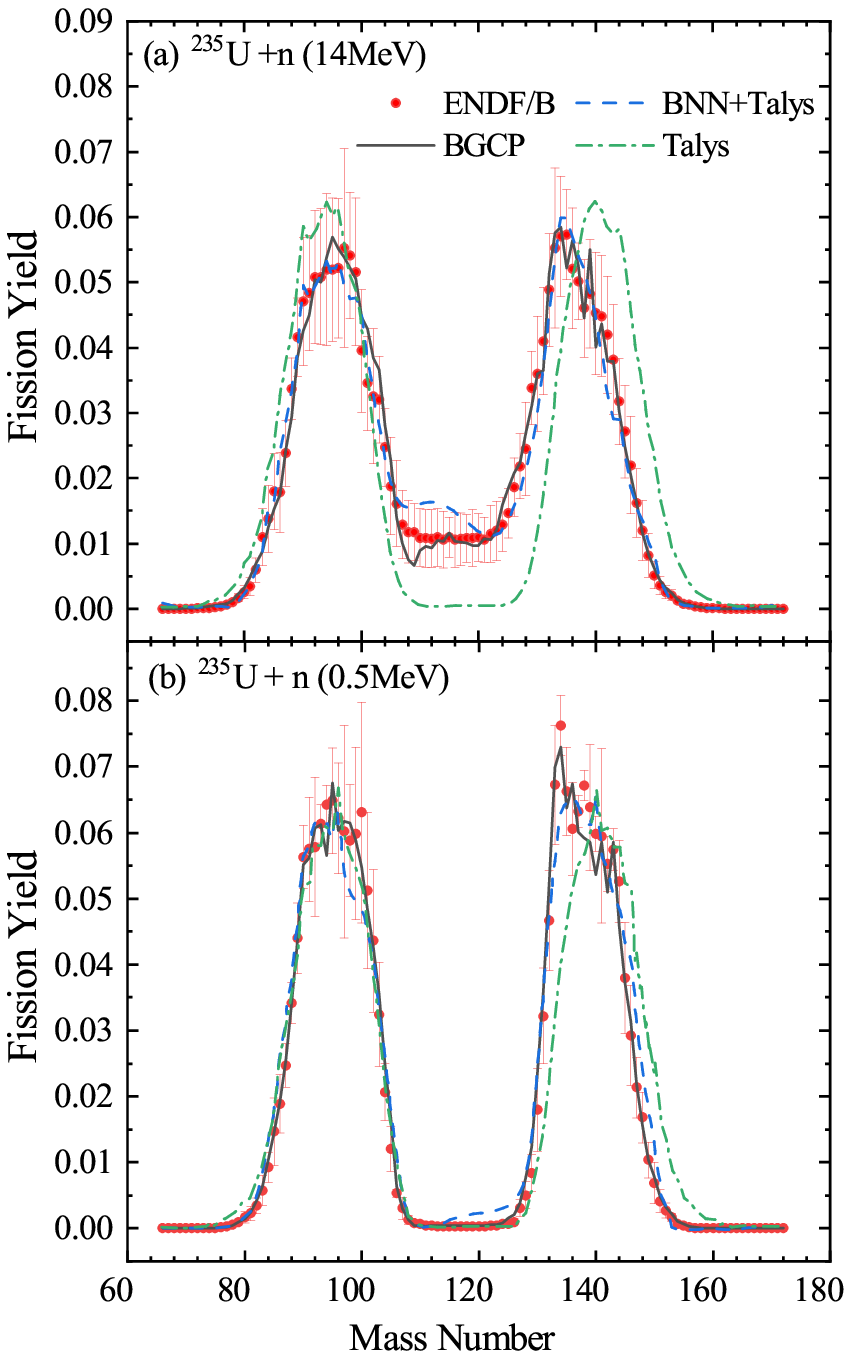}
	\caption{\label{U235}
		(Color online) $^{235}$U fission yield mass distribution predicted by various models and corresponding ENDF/B-VIII.0 data (red dot), at neutron incident energies of (a) 14 MeV and (b) 0.5 MeV. The black solid curves, blue dashed curves and dash-dotted curves represent the predictions of FYTD, BNN+TALYS and TALYS models respectively. BNN+TALYS and TALYS predictions are taken from Ref. \cite{wang_bayesian_2019}.
	}
\end{figure}

In this part, all $^{235}$U or $^{239}$Pu yield under the three energy points are set as missing values, and the FYTD model is used to reproduce the $^{235}$U or $^{239}$Pu yield data. Fig. \ref{U235} shows the predicted mass distribution of the products in $^{235}$U fission at 14 MeV and 0.5 MeV. In BNN+TALYS model, JENDL-4.0 data was used for model training, and this paper uses the ENDF/B-VIII.0 data. However, the fission yield data in JENDL database actually refers to that in ENDF/B database, the fission yield data in these two database are almost the same. In addition, the prediction by the BNN+TALYS model are also made when all $^{235}$U yield data in the learning set are removed. Thus, it is very suitable to compare with the prediction by the FYTD model. It can be seen from Fig. \ref{U235} that the TALYS model has a large deviation from the ENDF/B data at the heavy nucleus peak, and also fails to predict the neutron energy dependence of valley yields. After combining the trained BNN algorithm, the BNN+Talys model can correct these deviations. However, the BNN+Talys model still has defects. In the valley area (A=116$\thicksim$128) of Fig. \ref{U235}(b), the yield here is low, around $4\times10^{-4}$, but the BNN+TALYS model significantly overestimated this value, predicting a value around $2\times10^{-3}$. This phenomenon also appears in the area of A=104$\thicksim$120 in Fig. \ref{U235}(a). In general, the FYTD model performs better, it predicts the yields in the peak areas well, and successfully predicts the neutron energy dependence of the valley yields, only slightly underestimates the yields around A = 109. The accuracy of the FYTD model prediction can also be proved by $\chi_{N}^{2}$ in Table. \ref{chi2}.

This result proves that when all $^{235}$U yield data in the learning set are removed, the FYTD model can better capture information from other heavy nucleus yield data and predict the $^{235}$U yield data. At the same time, it can also capture the neutron energy dependence information of yield data and predict the variation trend of yield with neutron energy. Compared with algorithms such as BNN, the FYTD model preform better in fission yield evaluating and predicting.

\begin{table}[!htbp]
	\centering
	\caption{\label{chi2} The validation errors $\chi_{N}^{2}$ of various models. Errors for TALYS, BNN-40 and BNN-40 + TALYS are taken from Ref. \cite{wang_bayesian_2019}}
	\begin{tabular}{cc}
		\hline\hline
		Models & Validation $\chi_{N}^{2} (\times10^{-5})$  \\
		\hline
		TALYS \cite{koning_modern_2012} & 8.334  \\
		BNN-40 \cite{wang_bayesian_2019} & 1.640  \\
		BNN-40 + TALYS \cite{wang_bayesian_2019} & 1.134  \\
		FYTD & 0.701  \\
		\hline\hline
	\end{tabular}
\end{table}

\begin{figure*}
	\centering
	\includegraphics[width=16cm,angle=0]{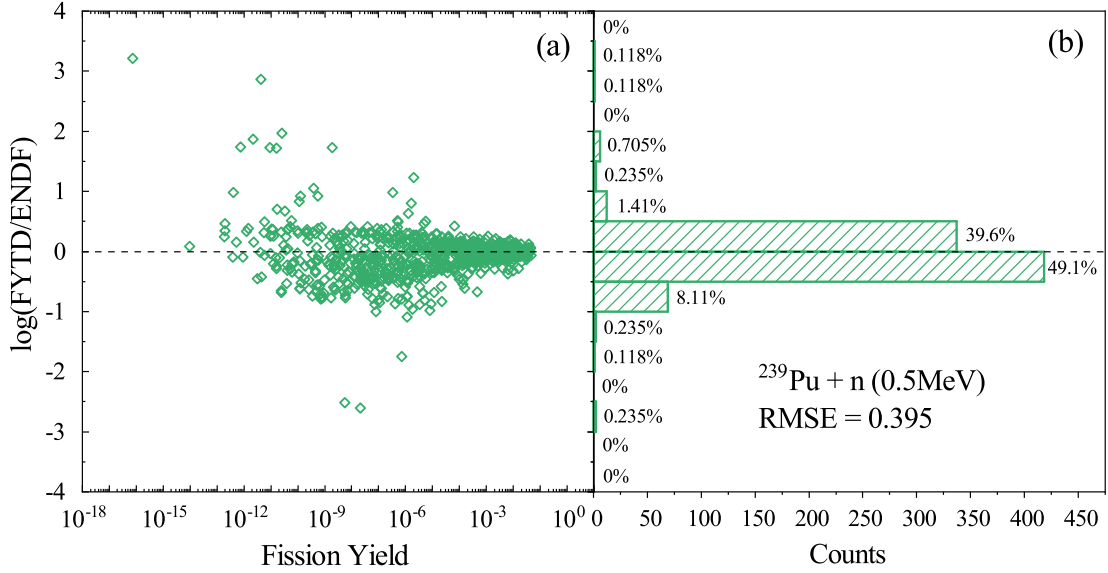}
	\caption{\label{Pu239}
		(Color online) Logarithm of the ratio of FYTD prediction to ENDF/B-VIII.0 data (logarithmic error) for 0.5 MeV neutron induced $^{239}$Pu fission. (a) is the distribution of errors and (b) is the count of errors of different sizes. Each dot represents a product, the ordinate is the prediction error of the product, and the abscissa is the yield of the product.
	}
\end{figure*}

Compared with the BNN and BNN+TALYS model, which only predicted the mass distribution of the yield, the FYTD model can also predict the isotopic distribution of the product. In order to more comprehensively show the difference between the predicted yield and the ENDF/B-VIII.0 data of 851 fission products, Fig. \ref{Pu239} shows the logarithmic error distribution of the fast neutron induced fission product yield of $^{239}$Pu and the RMSE of the predicted results. A dotted line with zero error is marked in the two figures. The closer the dot is to the line, the smaller the error. The RMSE of the prediction result at this time is 0.395, which is little higher than the 0.281 in Table. \ref{RMSE}. This is understandable, because the learning set of this prediction result removes all the data of all three energy points of $^{239}$Pu, while the test in Table. \ref{RMSE} only removes the data of one energy point. It can be seen from Fig. \ref{Pu239}(a) that the greater the yield value of the product, the higher the accuracy of the prediction. This confirms that the method of constructing tensors with partial logarithmic and partial linear coordinates can make nuclide prediction with large yield more accurate. Fig. \ref{Pu239}(b) shows the count of errors of different sizes. It can be seen that the logarithmic error of 98 $\%$ isotopes is within $\pm$ 1. that is, the difference between the predicted yield and ENDF/B data is within 1 order of magnitude. And for 88 $\%$ nuclides, the difference between the predicted yield and ENDF/B data is within 0.5 order of magnitude. However, it can be seen from Fig. \ref{Pu239}(a) that $^{239}$Pu fast neutron-induced fission yield data spans 16 orders of magnitude from $10^{-18}$ to $10^{-3}$, this prediction of isotope distribution by the FYTD model is relatively satisfactory.

\subsection{$^{238}$Np, $^{243}$Am and $^{236}$Np fission yield prediction}

After several verification, it is proved that the FYTD model has good learning and prediction ability for ENDF/B data. On this basis, the FYTD model is applied to the real data prediction to predict the missing data in the ENDF/B database. In order to verify whether the predicted results are reliable, the predicted yield data will be compared with JEFF database and experimental data.

\begin{figure}
	\centering
	\includegraphics[width=8.6cm,angle=0]{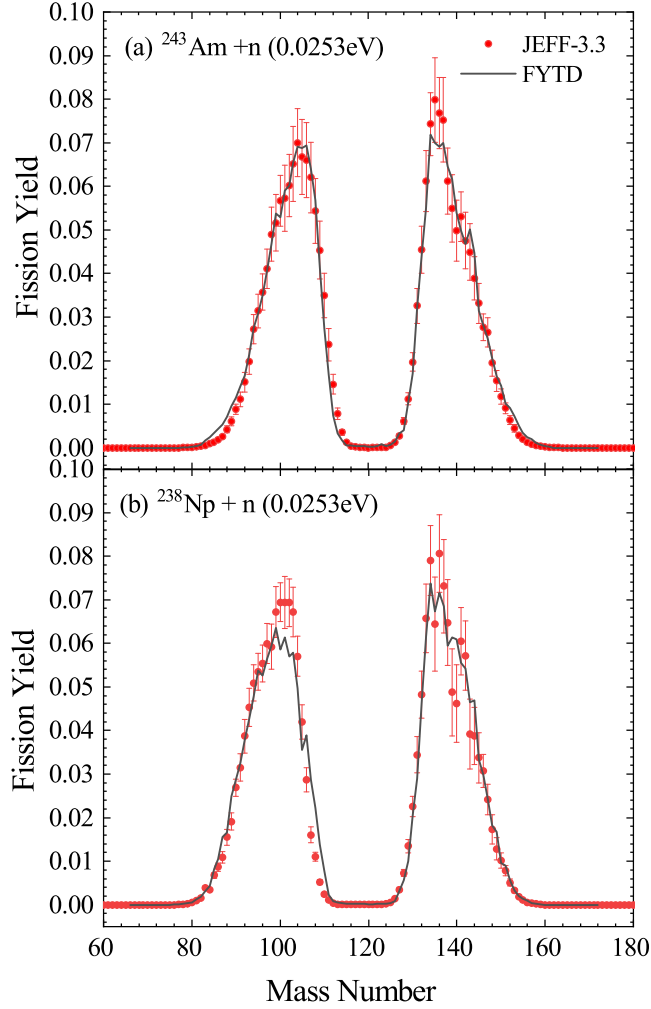}
	\caption{\label{JEFF}
		(Color online) Fission yield mass distribution predicted by FYTD model (black solid curves) and corresponding JEFF-3.3 data (red dots), for (a) $^{243}$Am + n at 0.0253 eV and (b) $^{238}$Np + n at 0.0253 eV.
	}
\end{figure}

Taking $^{243}$Am and $^{238}$Np as examples, their fission yield data under thermal neutron is missing in the ENDF/B-VIII.0 database, but they exists in the JEFF-3.3 database. The predicted fission yield will be compared with the data in JEFF-3.3. shows the comparison of yield mass distribution. It can be seen from Fig. \ref{JEFF}(a) that the prediction of $^{243}$Am by the FYTD model is basically within the error range of JEFF data. Only at the edge A=80$\thicksim$85, there is a certain deviation between the prediction results and the JEFF data. For prediction of $^{238}$Np in Fig. \ref{JEFF}(b), there is a certain difference of the light nucleus peak from JEFF data.

From the above two comparisons, it can be seen that the prediction of the FYTD model for the yield data of $^{243}$Am and $^{238}$Np under thermal neutron are consistent with the JEFF data, but this is based on the existing data. As can be seen from Table \ref{RMSE}, $^{243}$Am and $^{238}$Np have no data under thermal neutrons but under fast neutrons in ENDF/B database. It is relatively easy for the FYTD model to predict the data of a target nucleus at one energy points when the data at other energy point is known. However, the yield data of some target nuclei are very scarce, such as $^{236}$Np. Its independent yield data are missing in both ENDF/B and JEFF databases, and there are only a few experimental measurement data of chain yield. To predict the yield data of $^{236}$Np, it is necessary to capture information from other target nuclei, which tests the model's ability to learn and predict the dependence of target nucleus.

\begin{figure}
	\centering
	\includegraphics[width=8.6cm,angle=0]{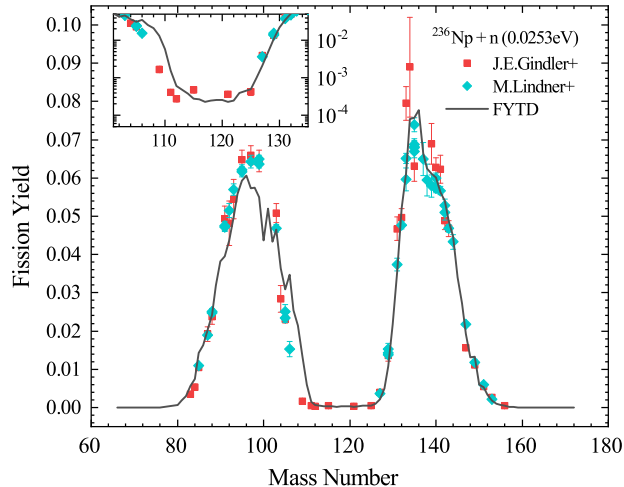}
	\caption{\label{Np236}
		(Color online) Fission yield mass distribution of $^{236}$Np + n at 0.0253 eV predicted by FYTD model (black solid curves) and two sets of corresponding experimental data (red and green dots) from Ref. \cite{lindner_reactor_1990, gindler_mass_nodate}. The inset panel on the upper left shows the valley areas in logarithmic scale.
	}
\end{figure}

Figrue \ref{Np236} show the mass distribution of fission products in thermal fission of $^{236}$Np measured by the two experimental groups and the corresponding predictions by the FYTD model. The panel on the upper left is added to show the valley areas in logarithmic scale. It can be seen that the predictions by the FYTD model at heavy nucleus peak agree with the two sets of experimental data, and there is a slight deviation near A = 100 at light nucleus peak.
As for the valley value, the predicted magnitude is consistent with the experimental data. This result proves the prediction ability of the FYTD model on the target nucleus dependence of fission yield.

\subsection{Prediction of 2 MeV neutron-induced fission yield}

The comparison and verification above proved the FYTD model ability to learn and predict neutron energy dependence and target nucleus dependence. However, there are few energy points in the yield data in the current database, a reliable model needs to be able to predict data at more energy points. And the above verification is also based on three energy points. In order to verify the scalability of the FYTD model over the incident neutron energy degrees of freedom, the only fission yield data under 2 MeV ($^{239}$Pu + n at 2 MeV) in ENDF/B-VIII.0 database is included in tensor construction. We test whether the model can predict the yield data of the remaining heavy nucleus at 2 MeV based on this yield data of $^{239}$Pu + n at 2 MeV.

\begin{figure}
	\centering
	\includegraphics[width=8.6cm,angle=0]{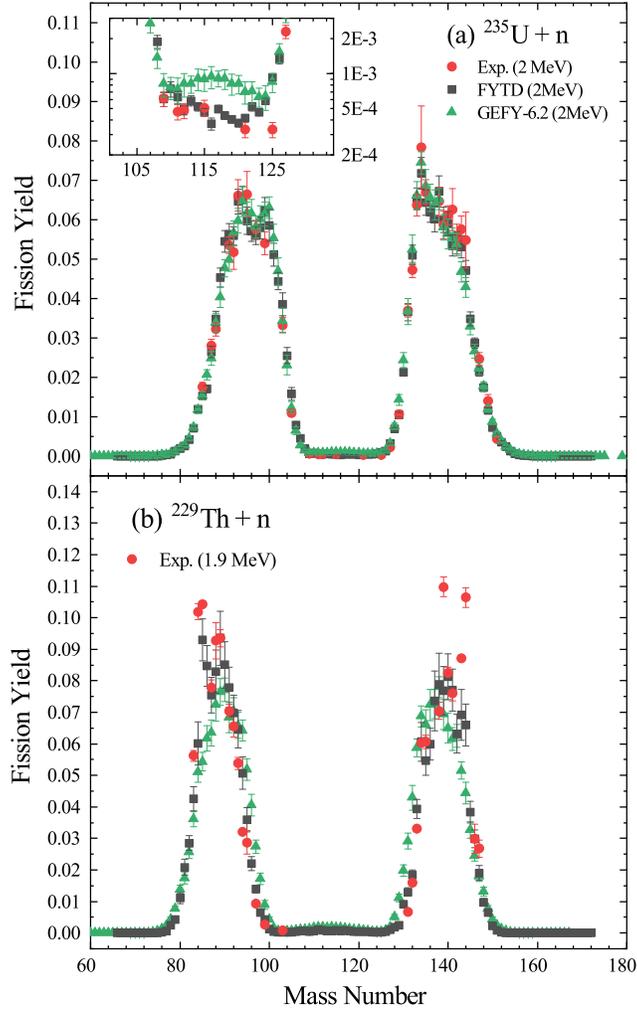}
	\caption{\label{2MeV}
		(Color online) Fission yield mass distribution for (a) $^{235}$U and (b) $^{229}$Th at 2 MeV predicted by FYTD model (black dots), GEF model (green dots) and corresponding experimental data (red dots) from Ref. \cite{glendenin_mass_1981,agarwal_mass_2008}. The inset panel shows the valley areas in logarithmic scale.
	}
\end{figure}

The predictions of yield at 2 MeV by the FYTD model are compared with the experimental data and the  predictions of the GEF model. The data of the GEF model comes from the database GEFY-6.2. Figrue \ref{2MeV} shows the mass distribution of the products in $^{235}$U and $^{229}$Th fissions at 2 MeV. It can be seen form Fig. \ref{2MeV}(a) that both predictions by the FYTD and GEF models agree well with experimental data in the peak area. In the valley area shown in the inset panel, the prediction by the FYTD model is more consistent with the experimental data than that by the GEF model. This may be due to the generally low yield in the valley area and the lack of experimental data in this area, resulting in a slightly insufficient predictive ability of traditional phenomenological models such as GEF.

For predictions for $^{229}$Th in Fig. \ref{2MeV}(b), no completely strict experimental data for $^{229}$Th at 2 MeV are published.
The experimental data at 1.9 MeV was used for comparison. According to energy dependence, it can be simply speculated that the data at 1.9 MeV should be slightly higher than the data at 2 MeV in the peak area, and lower in the valley area. The peak predicted by the FYTD model is just slightly lower than the experimental value and higher than that predicted by the GEF model. Overall, the predictions by the FYTD model are also more consistent with the experimental data than the predictions of the GEF model.

\begin{figure}
	\centering
	\includegraphics[width=8.6cm,angle=0]{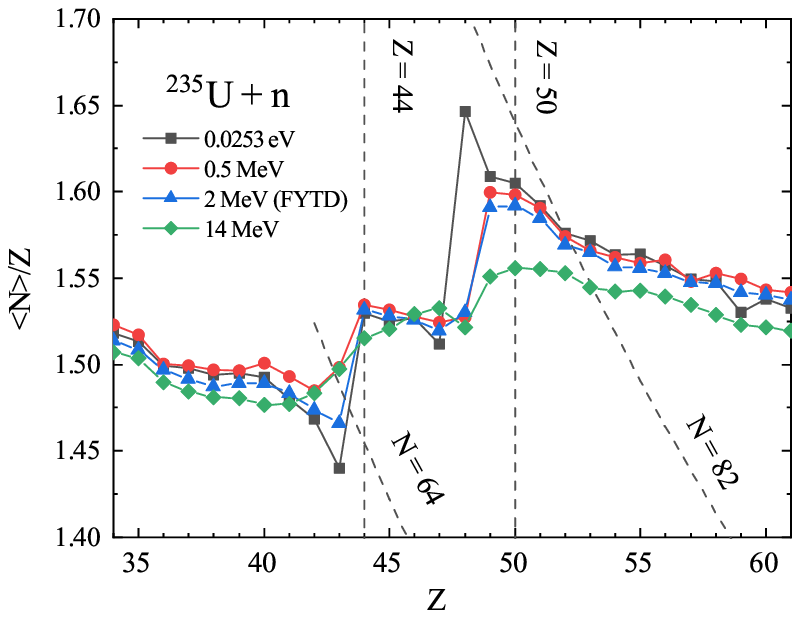}
	\caption{\label{excess}
		(Color online) Neutron excess as a function of product charge number for $^{235}$U + n fission at different energies, in which the 2 MeV results are predicted by the FYTD model. Spherical shell Z = 50, N =82 and deformed shells Z = 44, N = 64 are marked by dashed lines.
	}
\end{figure}

\begin{figure}
	\centering
	\includegraphics[width=8.6cm,angle=0]{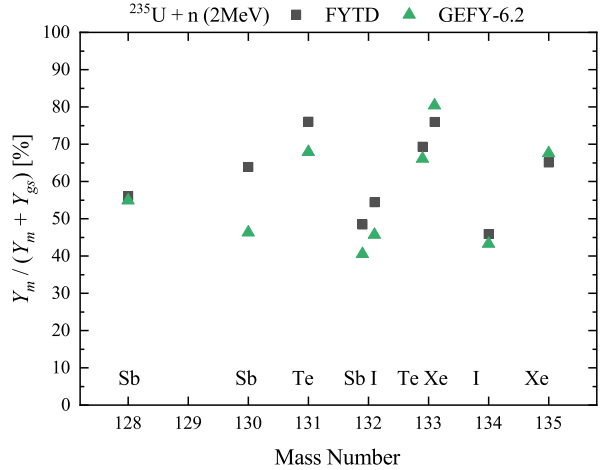}
	\caption{\label{isomeric}
		(Color online) Isomeric ratios predicted by FYTD model (black dots) and GEF model (green dots) for several fission products from $^{235}$U + n at 2 MeV.
	}
\end{figure}

Figures \ref{excess} and \ref{isomeric} verify the predicted fission yield induced by 2 MeV neutron of $^{235}$U from a more physical point of view. Figure \ref{excess} shows the neutron excess of $^{235}$U fission products induced by neutrons of different energies. The data of 2 MeV are the prediction by the FYTD model, and the rest are for ENDF/B data. The positions of some shells are marked in the figure. The definition of neutron excess is referred to the Ref. \cite{ramos_insight_2019} and is defined as the ratio of the average neutron number to the proton number of the product. This quantity can reflect the neutron proton composition of the fission product and the influence of shell effect on fission. From the data at 0.0253 eV, 0.5 MeV and 14 MeV in the figure, it can be observed that the neutron proton composition of the product is greatly affected by the shells of Z = 50, N = 82 and Z = 44, N = 64. However, with the increase of neutron incident energy, the excitation energy of fission system increases, and the curve gradually flattens. This is because the increase of excitation energy will hinder the influence of shell effect, and the increase of valley area in yield mass distribution is also caused by this reason. It can be seen from the figure that the predicted data under 2 MeV neutron incidence generally conforms to this law. In most areas, the predictions are between 0.5 MeV and 14 MeV data, and only deviate at Z = 43. This verification better proved that FYTD model learn and predict the impact of excitation energy on yield.

In the FYTD model, the isomeric states of products can be considered, and the corresponding data can be learned and predicted. Figure \ref{isomeric} shows the proportion of isomeric states for $^{128}$Sb, $^{130}$Sb, $^{131}$Te, $^{132}$Sb, $^{132}$I, $^{133}$Te, $^{133}$Xe, $^{134}$I and $^{135}$Xe produced in fission. It can be seen that except for $^{130}$Sb, the predictions by the FYTD model are consistent with those by the GEF model.
The above comparison proves the powerful ability of the multi-dimensional dependence learning, which can predict physical laws to a some extent.
Predictions for fission at 2 MeV also proves that the FYTD model is not only applicative at 0.0253 eV, 0.5 MeV and 14 MeV, but also extend to other neutron energy with the help of data for one target nucleus.
%If the FYTD model continues to mature and improve in the future, this may be helpful in the prediction and evaluation of fission yield data.

\section{\label{summary}CONCLUSION}

This work applied the tensor decomposition algorithm to the prediction of independent fission product yields. A Fission Yield Tensor Decomposition model was established and applied to a variety of yield predictions for verification.

The fission yields of $^{235}$U and $^{239}$Pu are set as missing values and then predicted. Compared with those by the Talys, BNN, and BNN+Talys models, the predictions by the FYTD model agree better with the ENDF/B-VIII.0 data. The predicted fission yields under fast neutron and high-energy neutron incident are compared. It is found that the FYTD model can reproduce the neutron energy dependence of the yield data. The data for $^{239}$Pu fission under fast neutron incident spans 16 orders of magnitude from 10$^{-3}$ to 10$^{-18}$. Comparison between the prediction and ENDF/B-VIII.0 data for 851 fission products shows that 98$\%$ of them agree with each other within 1 order of magnitude, and 88$\%$ within 0.5 order of magnitude.

By comparing the predictions with the data of $^{238}$Np and $^{243}$Am in the JEFF-3.3 database, as well as the experimental data of $^{236}$Np, the predictive ability of the model for the target nucleus dependence of the yield data is demonstrated. The scalability of the yield tensor decomposition model over the incident neutron energy degrees of freedom is examined. After adding a set of 2 MeV neutron-induced $^{239}$Pu fission yield data into the yield tensor, the 2 MeV neutron-induced fission yields of $^{235}$U and $^{229}$Th are predicted. The comparison with the experimental data shows that the prediction of this work is similar to the prediction by the GEF model in the peak area and more accurate in the valley area. At the same time, the prediction of the ratio of isomeric states and excess of neutrons in the product also proves that the model has the ability to learn and predict the laws of physics.

The multiple rounds of comparative verification show that the FYTD model can capture the complex multi-dimensional dependence of fission yields and make reasonable predictions. At the same time, the model has scalability in the energy dimension, and also has a certain ability to predict physical laws. Facing the demands of various nuclear energy devices and systems with complex neutron spectrum in the future, the scalability of the FYTD model in the energy dimension makes it potentially valuable for future application.
%For the future application, the following aspects should be considered.
%This is the first time tensor decomposition has been applied to the prediction of fission yields, there are still many shortcoming in FYTD model at present.
%There is a lack of physical images in the model, and the error range of prediction is too small, which can only be observed in Fig. \ref{2MeV}.
The predictions and evaluations of this work are based on the ENDF/B-VIII.0 database.
In the future we will conduct evaluations based on the EXFOR database.
%If this kind of FYTD models continues to mature and improve in the future, tensor decomposition may become a powerful tool for evaluating and predicting complex nuclear data to help nuclear physics research.

\section*{ACKNOWLEDGMENTS}

This work was supported by the National Natural Science Foundation of China under Grants No. 11875328, 12075327 and 12105170.

%\begin{thebibliography}{10}
%\bibitem{zpa119} D. Boilley, Y. Abe, S. Ayik, and E. Suraud, Z. Phys. A 349, 119 (1994).
%\end{thebibliography}

% BibTeX users please use one of
%\bibliographystyle{spbasic}      % basic style, author-year citations
%\bibliographystyle{spmpsci}      % mathematics and physical sciences
       % APS-like style for physics
%\bibliography{}   % name your BibTeX data base
%\bibliographystyle{spphys}
\bibliography{mybibfile}

%merlin.mbs apsrev4-1.bst 2010-07-25 4.21a (PWD, AO, DPC) hacked
%Control: key (0)
%Control: author (8) initials jnrlst
%Control: editor formatted (1) identically to author
%Control: production of article title (-1) disabled
%Control: page (0) single
%Control: year (1) truncated
%Control: production of eprint (0) enabled
\begin{thebibliography}{43}%
\makeatletter
\providecommand \@ifxundefined [1]{%
 \@ifx{#1\undefined}
}%
\providecommand \@ifnum [1]{%
 \ifnum #1\expandafter \@firstoftwo
 \else \expandafter \@secondoftwo
 \fi
}%
\providecommand \@ifx [1]{%
 \ifx #1\expandafter \@firstoftwo
 \else \expandafter \@secondoftwo
 \fi
}%
\providecommand \natexlab [1]{#1}%
\providecommand \enquote  [1]{``#1''}%
\providecommand \bibnamefont  [1]{#1}%
\providecommand \bibfnamefont [1]{#1}%
\providecommand \citenamefont [1]{#1}%
\providecommand \href@noop [0]{\@secondoftwo}%
\providecommand \href [0]{\begingroup \@sanitize@url \@href}%
\providecommand \@href[1]{\@@startlink{#1}\@@href}%
\providecommand \@@href[1]{\endgroup#1\@@endlink}%
\providecommand \@sanitize@url [0]{\catcode `\\12\catcode `\$12\catcode
  `\&12\catcode `\#12\catcode `\^12\catcode `\_12\catcode `\%12\relax}%
\providecommand \@@startlink[1]{}%
\providecommand \@@endlink[0]{}%
\providecommand \url  [0]{\begingroup\@sanitize@url \@url }%
\providecommand \@url [1]{\endgroup\@href {#1}{\urlprefix }}%
\providecommand \urlprefix  [0]{URL }%
\providecommand \Eprint [0]{\href }%
\providecommand \doibase [0]{http://dx.doi.org/}%
\providecommand \selectlanguage [0]{\@gobble}%
\providecommand \bibinfo  [0]{\@secondoftwo}%
\providecommand \bibfield  [0]{\@secondoftwo}%
\providecommand \translation [1]{[#1]}%
\providecommand \BibitemOpen [0]{}%
\providecommand \bibitemStop [0]{}%
\providecommand \bibitemNoStop [0]{.\EOS\space}%
\providecommand \EOS [0]{\spacefactor3000\relax}%
\providecommand \BibitemShut  [1]{\csname bibitem#1\endcsname}%
\let\auto@bib@innerbib\@empty
%</preamble>
\bibitem [{\citenamefont {Hahn}\ and\ \citenamefont
  {Strassmann}(1939)}]{hahn1939nachweis}%
  \BibitemOpen
  \bibfield  {author} {\bibinfo {author} {\bibfnamefont {O.}~\bibnamefont
  {Hahn}}\ and\ \bibinfo {author} {\bibfnamefont {F.}~\bibnamefont
  {Strassmann}},\ }\href@noop {} {\bibfield  {journal} {\bibinfo  {journal}
  {Naturwissenschaften}\ }\textbf {\bibinfo {volume} {27}},\ \bibinfo {pages}
  {11} (\bibinfo {year} {1939})}\BibitemShut {NoStop}%
\bibitem [{\citenamefont {Meitner}\ and\ \citenamefont
  {Frisch}(1939)}]{meitner1939disintegration}%
  \BibitemOpen
  \bibfield  {author} {\bibinfo {author} {\bibfnamefont {L.}~\bibnamefont
  {Meitner}}\ and\ \bibinfo {author} {\bibfnamefont {O.~R.}\ \bibnamefont
  {Frisch}},\ }\href@noop {} {\bibfield  {journal} {\bibinfo  {journal}
  {Nature}\ }\textbf {\bibinfo {volume} {143}},\ \bibinfo {pages} {239}
  (\bibinfo {year} {1939})}\BibitemShut {NoStop}%
\bibitem [{\citenamefont {Schunck}\ and\ \citenamefont
  {Robledo}(2016)}]{schunck2016microscopic}%
  \BibitemOpen
  \bibfield  {author} {\bibinfo {author} {\bibfnamefont {N.}~\bibnamefont
  {Schunck}}\ and\ \bibinfo {author} {\bibfnamefont {L.}~\bibnamefont
  {Robledo}},\ }\href@noop {} {\bibfield  {journal} {\bibinfo  {journal}
  {Reports on Progress in Physics}\ }\textbf {\bibinfo {volume} {79}},\
  \bibinfo {pages} {116301} (\bibinfo {year} {2016})}\BibitemShut {NoStop}%
\bibitem [{\citenamefont {Bender}\ \emph {et~al.}(2020)\citenamefont {Bender},
  \citenamefont {Bernard}, \citenamefont {Bertsch}, \citenamefont {Chiba},
  \citenamefont {Dobaczewski}, \citenamefont {Dubray}, \citenamefont
  {Giuliani}, \citenamefont {Hagino}, \citenamefont {Lacroix}, \citenamefont
  {Li} \emph {et~al.}}]{bender_future_2020}%
  \BibitemOpen
  \bibfield  {author} {\bibinfo {author} {\bibfnamefont {M.}~\bibnamefont
  {Bender}}, \bibinfo {author} {\bibfnamefont {R.}~\bibnamefont {Bernard}},
  \bibinfo {author} {\bibfnamefont {G.}~\bibnamefont {Bertsch}}, \bibinfo
  {author} {\bibfnamefont {S.}~\bibnamefont {Chiba}}, \bibinfo {author}
  {\bibfnamefont {J.}~\bibnamefont {Dobaczewski}}, \bibinfo {author}
  {\bibfnamefont {N.}~\bibnamefont {Dubray}}, \bibinfo {author} {\bibfnamefont
  {S.~A.}\ \bibnamefont {Giuliani}}, \bibinfo {author} {\bibfnamefont
  {K.}~\bibnamefont {Hagino}}, \bibinfo {author} {\bibfnamefont
  {D.}~\bibnamefont {Lacroix}}, \bibinfo {author} {\bibfnamefont
  {Z.}~\bibnamefont {Li}},  \emph {et~al.},\ }\href@noop {} {\bibfield
  {journal} {\bibinfo  {journal} {Journal of Physics G: Nuclear and Particle
  Physics}\ }\textbf {\bibinfo {volume} {47}},\ \bibinfo {pages} {113002}
  (\bibinfo {year} {2020})}\BibitemShut {NoStop}%
\bibitem [{\citenamefont {Hamilton}\ \emph {et~al.}(2013)\citenamefont
  {Hamilton}, \citenamefont {Hofmann},\ and\ \citenamefont
  {Oganessian}}]{hamilton_search_2013}%
  \BibitemOpen
  \bibfield  {author} {\bibinfo {author} {\bibfnamefont {J.}~\bibnamefont
  {Hamilton}}, \bibinfo {author} {\bibfnamefont {S.}~\bibnamefont {Hofmann}}, \
  and\ \bibinfo {author} {\bibfnamefont {Y.}~\bibnamefont {Oganessian}},\
  }\href {\doibase 10.1146/annurev-nucl-102912-144535} {\bibfield  {journal}
  {\bibinfo  {journal} {Annual Review of Nuclear and Particle Science}\
  }\textbf {\bibinfo {volume} {63}},\ \bibinfo {pages} {383} (\bibinfo {year}
  {2013})}\BibitemShut {NoStop}%
\bibitem [{\citenamefont {Pei}\ \emph {et~al.}(2009)\citenamefont {Pei},
  \citenamefont {Nazarewicz}, \citenamefont {Sheikh},\ and\ \citenamefont
  {Kerman}}]{pei_fission_2009}%
  \BibitemOpen
  \bibfield  {author} {\bibinfo {author} {\bibfnamefont {J.~C.}\ \bibnamefont
  {Pei}}, \bibinfo {author} {\bibfnamefont {W.}~\bibnamefont {Nazarewicz}},
  \bibinfo {author} {\bibfnamefont {J.~A.}\ \bibnamefont {Sheikh}}, \ and\
  \bibinfo {author} {\bibfnamefont {A.~K.}\ \bibnamefont {Kerman}},\ }\href
  {\doibase 10.1103/PhysRevLett.102.192501} {\bibfield  {journal} {\bibinfo
  {journal} {Physical Review Letters}\ }\textbf {\bibinfo {volume} {102}},\
  \bibinfo {pages} {192501} (\bibinfo {year} {2009})}\BibitemShut {NoStop}%
\bibitem [{\citenamefont {Eichler}\ \emph {et~al.}(2015)\citenamefont
  {Eichler}, \citenamefont {Arcones}, \citenamefont {Kelic}, \citenamefont
  {Korobkin}, \citenamefont {Langanke}, \citenamefont {Marketin}, \citenamefont
  {Martinez-Pinedo}, \citenamefont {Panov}, \citenamefont {Rauscher},
  \citenamefont {Rosswog}, \citenamefont {Winteler}, \citenamefont {Zinner},\
  and\ \citenamefont {Thielemann}}]{eichler_role_2015}%
  \BibitemOpen
  \bibfield  {author} {\bibinfo {author} {\bibfnamefont {M.}~\bibnamefont
  {Eichler}}, \bibinfo {author} {\bibfnamefont {A.}~\bibnamefont {Arcones}},
  \bibinfo {author} {\bibfnamefont {A.}~\bibnamefont {Kelic}}, \bibinfo
  {author} {\bibfnamefont {O.}~\bibnamefont {Korobkin}}, \bibinfo {author}
  {\bibfnamefont {K.}~\bibnamefont {Langanke}}, \bibinfo {author}
  {\bibfnamefont {T.}~\bibnamefont {Marketin}}, \bibinfo {author}
  {\bibfnamefont {G.}~\bibnamefont {Martinez-Pinedo}}, \bibinfo {author}
  {\bibfnamefont {I.}~\bibnamefont {Panov}}, \bibinfo {author} {\bibfnamefont
  {T.}~\bibnamefont {Rauscher}}, \bibinfo {author} {\bibfnamefont
  {S.}~\bibnamefont {Rosswog}}, \bibinfo {author} {\bibfnamefont
  {C.}~\bibnamefont {Winteler}}, \bibinfo {author} {\bibfnamefont {N.~T.}\
  \bibnamefont {Zinner}}, \ and\ \bibinfo {author} {\bibfnamefont {F.-K.}\
  \bibnamefont {Thielemann}},\ }\href {\doibase 10.1088/0004-637X/808/1/30}
  {\bibfield  {journal} {\bibinfo  {journal} {The Astrophysical Journal}\
  }\textbf {\bibinfo {volume} {808}},\ \bibinfo {pages} {30} (\bibinfo {year}
  {2015})}\BibitemShut {NoStop}%
\bibitem [{\citenamefont {Mueller}\ \emph {et~al.}(2011)\citenamefont
  {Mueller}, \citenamefont {Lhuillier}, \citenamefont {Fallot}, \citenamefont
  {Letourneau}, \citenamefont {Cormon}, \citenamefont {Fechner}, \citenamefont
  {Giot}, \citenamefont {Lasserre}, \citenamefont {Martino}, \citenamefont
  {Mention}, \citenamefont {Porta},\ and\ \citenamefont
  {Yermia}}]{mueller_improved_2011}%
  \BibitemOpen
  \bibfield  {author} {\bibinfo {author} {\bibfnamefont {T.~A.}\ \bibnamefont
  {Mueller}}, \bibinfo {author} {\bibfnamefont {D.}~\bibnamefont {Lhuillier}},
  \bibinfo {author} {\bibfnamefont {M.}~\bibnamefont {Fallot}}, \bibinfo
  {author} {\bibfnamefont {A.}~\bibnamefont {Letourneau}}, \bibinfo {author}
  {\bibfnamefont {S.}~\bibnamefont {Cormon}}, \bibinfo {author} {\bibfnamefont
  {M.}~\bibnamefont {Fechner}}, \bibinfo {author} {\bibfnamefont
  {L.}~\bibnamefont {Giot}}, \bibinfo {author} {\bibfnamefont {T.}~\bibnamefont
  {Lasserre}}, \bibinfo {author} {\bibfnamefont {J.}~\bibnamefont {Martino}},
  \bibinfo {author} {\bibfnamefont {G.}~\bibnamefont {Mention}}, \bibinfo
  {author} {\bibfnamefont {A.}~\bibnamefont {Porta}}, \ and\ \bibinfo {author}
  {\bibfnamefont {F.}~\bibnamefont {Yermia}},\ }\href {\doibase
  10.1103/PhysRevC.83.054615} {\bibfield  {journal} {\bibinfo  {journal}
  {Physical Review C}\ }\textbf {\bibinfo {volume} {83}},\ \bibinfo {pages}
  {054615} (\bibinfo {year} {2011})}\BibitemShut {NoStop}%
\bibitem [{\citenamefont {Bernstein}\ \emph {et~al.}(2019)\citenamefont
  {Bernstein}, \citenamefont {Brown}, \citenamefont {Koning}, \citenamefont
  {Rearden}, \citenamefont {Romano}, \citenamefont {Sonzogni}, \citenamefont
  {Voyles},\ and\ \citenamefont {Younes}}]{bernstein_our_2019}%
  \BibitemOpen
  \bibfield  {author} {\bibinfo {author} {\bibfnamefont {L.~A.}\ \bibnamefont
  {Bernstein}}, \bibinfo {author} {\bibfnamefont {D.~A.}\ \bibnamefont
  {Brown}}, \bibinfo {author} {\bibfnamefont {A.~J.}\ \bibnamefont {Koning}},
  \bibinfo {author} {\bibfnamefont {B.~T.}\ \bibnamefont {Rearden}}, \bibinfo
  {author} {\bibfnamefont {C.~E.}\ \bibnamefont {Romano}}, \bibinfo {author}
  {\bibfnamefont {A.~A.}\ \bibnamefont {Sonzogni}}, \bibinfo {author}
  {\bibfnamefont {A.~S.}\ \bibnamefont {Voyles}}, \ and\ \bibinfo {author}
  {\bibfnamefont {W.}~\bibnamefont {Younes}},\ }\href {\doibase
  10.1146/annurev-nucl-101918-023708} {\bibfield  {journal} {\bibinfo
  {journal} {Annual Review of Nuclear and Particle Science}\ }\textbf {\bibinfo
  {volume} {69}},\ \bibinfo {pages} {109} (\bibinfo {year} {2019})}\BibitemShut
  {NoStop}%
\bibitem [{\citenamefont {Ramos}\ \emph {et~al.}(2019)\citenamefont {Ramos},
  \citenamefont {Caamaño}, \citenamefont {Farget}, \citenamefont
  {Rodríguez-Tajes}, \citenamefont {Audouin}, \citenamefont {Benlliure},
  \citenamefont {Casarejos}, \citenamefont {Clement}, \citenamefont {Cortina},
  \citenamefont {Delaune}, \citenamefont {Derkx}, \citenamefont {Dijon},
  \citenamefont {Doré}, \citenamefont {Fernández-Domínguez}, \citenamefont
  {de~France}, \citenamefont {Heinz}, \citenamefont {Jacquot}, \citenamefont
  {Paradela}, \citenamefont {Rejmund}, \citenamefont {Roger}, \citenamefont
  {Salsac},\ and\ \citenamefont {Schmitt}}]{ramos_insight_2019}%
  \BibitemOpen
  \bibfield  {author} {\bibinfo {author} {\bibfnamefont {D.}~\bibnamefont
  {Ramos}}, \bibinfo {author} {\bibfnamefont {M.}~\bibnamefont {Caamaño}},
  \bibinfo {author} {\bibfnamefont {F.}~\bibnamefont {Farget}}, \bibinfo
  {author} {\bibfnamefont {C.}~\bibnamefont {Rodríguez-Tajes}}, \bibinfo
  {author} {\bibfnamefont {L.}~\bibnamefont {Audouin}}, \bibinfo {author}
  {\bibfnamefont {J.}~\bibnamefont {Benlliure}}, \bibinfo {author}
  {\bibfnamefont {E.}~\bibnamefont {Casarejos}}, \bibinfo {author}
  {\bibfnamefont {E.}~\bibnamefont {Clement}}, \bibinfo {author} {\bibfnamefont
  {D.}~\bibnamefont {Cortina}}, \bibinfo {author} {\bibfnamefont
  {O.}~\bibnamefont {Delaune}}, \bibinfo {author} {\bibfnamefont
  {X.}~\bibnamefont {Derkx}}, \bibinfo {author} {\bibfnamefont
  {A.}~\bibnamefont {Dijon}}, \bibinfo {author} {\bibfnamefont
  {D.}~\bibnamefont {Doré}}, \bibinfo {author} {\bibfnamefont
  {B.}~\bibnamefont {Fernández-Domínguez}}, \bibinfo {author} {\bibfnamefont
  {G.}~\bibnamefont {de~France}}, \bibinfo {author} {\bibfnamefont
  {A.}~\bibnamefont {Heinz}}, \bibinfo {author} {\bibfnamefont
  {B.}~\bibnamefont {Jacquot}}, \bibinfo {author} {\bibfnamefont
  {C.}~\bibnamefont {Paradela}}, \bibinfo {author} {\bibfnamefont
  {M.}~\bibnamefont {Rejmund}}, \bibinfo {author} {\bibfnamefont
  {T.}~\bibnamefont {Roger}}, \bibinfo {author} {\bibfnamefont {M.-D.}\
  \bibnamefont {Salsac}}, \ and\ \bibinfo {author} {\bibfnamefont
  {C.}~\bibnamefont {Schmitt}},\ }\href {\doibase 10.1103/PhysRevC.99.024615}
  {\bibfield  {journal} {\bibinfo  {journal} {Physical Review C}\ }\textbf
  {\bibinfo {volume} {99}},\ \bibinfo {pages} {024615} (\bibinfo {year}
  {2019})}\BibitemShut {NoStop}%
\bibitem [{\citenamefont {Denschlag}(1986)}]{denschlag_independent_1986}%
  \BibitemOpen
  \bibfield  {author} {\bibinfo {author} {\bibfnamefont {H.~O.}\ \bibnamefont
  {Denschlag}},\ }\href {\doibase 10.13182/NSE86-A18345} {\bibfield  {journal}
  {\bibinfo  {journal} {Nuclear Science and Engineering}\ }\textbf {\bibinfo
  {volume} {94}},\ \bibinfo {pages} {337} (\bibinfo {year} {1986})}\BibitemShut
  {NoStop}%
\bibitem [{\citenamefont {Brown}\ \emph {et~al.}(2018)\citenamefont {Brown},
  \citenamefont {Chadwick}, \citenamefont {Capote}, \citenamefont {Kahler},
  \citenamefont {Trkov}, \citenamefont {Herman}, \citenamefont {Sonzogni},
  \citenamefont {Danon}, \citenamefont {Carlson}, \citenamefont {Dunn},
  \citenamefont {Smith}, \citenamefont {Hale}, \citenamefont {Arbanas},
  \citenamefont {Arcilla}, \citenamefont {Bates}, \citenamefont {Beck},
  \citenamefont {Becker}, \citenamefont {Brown}, \citenamefont {Casperson},
  \citenamefont {Conlin}, \citenamefont {Cullen}, \citenamefont {Descalle},
  \citenamefont {Firestone}, \citenamefont {Gaines}, \citenamefont {Guber},
  \citenamefont {Hawari}, \citenamefont {Holmes}, \citenamefont {Johnson},
  \citenamefont {Kawano}, \citenamefont {Kiedrowski}, \citenamefont {Koning},
  \citenamefont {Kopecky}, \citenamefont {Leal}, \citenamefont {Lestone},
  \citenamefont {Lubitz}, \citenamefont {Márquez~Damián}, \citenamefont
  {Mattoon}, \citenamefont {McCutchan}, \citenamefont {Mughabghab},
  \citenamefont {Navratil}, \citenamefont {Neudecker}, \citenamefont {Nobre},
  \citenamefont {Noguere}, \citenamefont {Paris}, \citenamefont {Pigni},
  \citenamefont {Plompen}, \citenamefont {Pritychenko}, \citenamefont
  {Pronyaev}, \citenamefont {Roubtsov}, \citenamefont {Rochman}, \citenamefont
  {Romano}, \citenamefont {Schillebeeckx}, \citenamefont {Simakov},
  \citenamefont {Sin}, \citenamefont {Sirakov}, \citenamefont {Sleaford},
  \citenamefont {Sobes}, \citenamefont {Soukhovitskii}, \citenamefont {Stetcu},
  \citenamefont {Talou}, \citenamefont {Thompson}, \citenamefont {van~der
  Marck}, \citenamefont {Welser-Sherrill}, \citenamefont {Wiarda},
  \citenamefont {White}, \citenamefont {Wormald}, \citenamefont {Wright},
  \citenamefont {Zerkle}, \citenamefont {Žerovnik},\ and\ \citenamefont
  {Zhu}}]{brown_endfb-viii0_2018}%
  \BibitemOpen
  \bibfield  {author} {\bibinfo {author} {\bibfnamefont {D.}~\bibnamefont
  {Brown}}, \bibinfo {author} {\bibfnamefont {M.}~\bibnamefont {Chadwick}},
  \bibinfo {author} {\bibfnamefont {R.}~\bibnamefont {Capote}}, \bibinfo
  {author} {\bibfnamefont {A.}~\bibnamefont {Kahler}}, \bibinfo {author}
  {\bibfnamefont {A.}~\bibnamefont {Trkov}}, \bibinfo {author} {\bibfnamefont
  {M.}~\bibnamefont {Herman}}, \bibinfo {author} {\bibfnamefont
  {A.}~\bibnamefont {Sonzogni}}, \bibinfo {author} {\bibfnamefont
  {Y.}~\bibnamefont {Danon}}, \bibinfo {author} {\bibfnamefont
  {A.}~\bibnamefont {Carlson}}, \bibinfo {author} {\bibfnamefont
  {M.}~\bibnamefont {Dunn}}, \bibinfo {author} {\bibfnamefont {D.}~\bibnamefont
  {Smith}}, \bibinfo {author} {\bibfnamefont {G.}~\bibnamefont {Hale}},
  \bibinfo {author} {\bibfnamefont {G.}~\bibnamefont {Arbanas}}, \bibinfo
  {author} {\bibfnamefont {R.}~\bibnamefont {Arcilla}}, \bibinfo {author}
  {\bibfnamefont {C.}~\bibnamefont {Bates}}, \bibinfo {author} {\bibfnamefont
  {B.}~\bibnamefont {Beck}}, \bibinfo {author} {\bibfnamefont {B.}~\bibnamefont
  {Becker}}, \bibinfo {author} {\bibfnamefont {F.}~\bibnamefont {Brown}},
  \bibinfo {author} {\bibfnamefont {R.}~\bibnamefont {Casperson}}, \bibinfo
  {author} {\bibfnamefont {J.}~\bibnamefont {Conlin}}, \bibinfo {author}
  {\bibfnamefont {D.}~\bibnamefont {Cullen}}, \bibinfo {author} {\bibfnamefont
  {M.-A.}\ \bibnamefont {Descalle}}, \bibinfo {author} {\bibfnamefont
  {R.}~\bibnamefont {Firestone}}, \bibinfo {author} {\bibfnamefont
  {T.}~\bibnamefont {Gaines}}, \bibinfo {author} {\bibfnamefont
  {K.}~\bibnamefont {Guber}}, \bibinfo {author} {\bibfnamefont
  {A.}~\bibnamefont {Hawari}}, \bibinfo {author} {\bibfnamefont
  {J.}~\bibnamefont {Holmes}}, \bibinfo {author} {\bibfnamefont
  {T.}~\bibnamefont {Johnson}}, \bibinfo {author} {\bibfnamefont
  {T.}~\bibnamefont {Kawano}}, \bibinfo {author} {\bibfnamefont
  {B.}~\bibnamefont {Kiedrowski}}, \bibinfo {author} {\bibfnamefont
  {A.}~\bibnamefont {Koning}}, \bibinfo {author} {\bibfnamefont
  {S.}~\bibnamefont {Kopecky}}, \bibinfo {author} {\bibfnamefont
  {L.}~\bibnamefont {Leal}}, \bibinfo {author} {\bibfnamefont {J.}~\bibnamefont
  {Lestone}}, \bibinfo {author} {\bibfnamefont {C.}~\bibnamefont {Lubitz}},
  \bibinfo {author} {\bibfnamefont {J.}~\bibnamefont {Márquez~Damián}},
  \bibinfo {author} {\bibfnamefont {C.}~\bibnamefont {Mattoon}}, \bibinfo
  {author} {\bibfnamefont {E.}~\bibnamefont {McCutchan}}, \bibinfo {author}
  {\bibfnamefont {S.}~\bibnamefont {Mughabghab}}, \bibinfo {author}
  {\bibfnamefont {P.}~\bibnamefont {Navratil}}, \bibinfo {author}
  {\bibfnamefont {D.}~\bibnamefont {Neudecker}}, \bibinfo {author}
  {\bibfnamefont {G.}~\bibnamefont {Nobre}}, \bibinfo {author} {\bibfnamefont
  {G.}~\bibnamefont {Noguere}}, \bibinfo {author} {\bibfnamefont
  {M.}~\bibnamefont {Paris}}, \bibinfo {author} {\bibfnamefont
  {M.}~\bibnamefont {Pigni}}, \bibinfo {author} {\bibfnamefont
  {A.}~\bibnamefont {Plompen}}, \bibinfo {author} {\bibfnamefont
  {B.}~\bibnamefont {Pritychenko}}, \bibinfo {author} {\bibfnamefont
  {V.}~\bibnamefont {Pronyaev}}, \bibinfo {author} {\bibfnamefont
  {D.}~\bibnamefont {Roubtsov}}, \bibinfo {author} {\bibfnamefont
  {D.}~\bibnamefont {Rochman}}, \bibinfo {author} {\bibfnamefont
  {P.}~\bibnamefont {Romano}}, \bibinfo {author} {\bibfnamefont
  {P.}~\bibnamefont {Schillebeeckx}}, \bibinfo {author} {\bibfnamefont
  {S.}~\bibnamefont {Simakov}}, \bibinfo {author} {\bibfnamefont
  {M.}~\bibnamefont {Sin}}, \bibinfo {author} {\bibfnamefont {I.}~\bibnamefont
  {Sirakov}}, \bibinfo {author} {\bibfnamefont {B.}~\bibnamefont {Sleaford}},
  \bibinfo {author} {\bibfnamefont {V.}~\bibnamefont {Sobes}}, \bibinfo
  {author} {\bibfnamefont {E.}~\bibnamefont {Soukhovitskii}}, \bibinfo {author}
  {\bibfnamefont {I.}~\bibnamefont {Stetcu}}, \bibinfo {author} {\bibfnamefont
  {P.}~\bibnamefont {Talou}}, \bibinfo {author} {\bibfnamefont
  {I.}~\bibnamefont {Thompson}}, \bibinfo {author} {\bibfnamefont
  {S.}~\bibnamefont {van~der Marck}}, \bibinfo {author} {\bibfnamefont
  {L.}~\bibnamefont {Welser-Sherrill}}, \bibinfo {author} {\bibfnamefont
  {D.}~\bibnamefont {Wiarda}}, \bibinfo {author} {\bibfnamefont
  {M.}~\bibnamefont {White}}, \bibinfo {author} {\bibfnamefont
  {J.}~\bibnamefont {Wormald}}, \bibinfo {author} {\bibfnamefont
  {R.}~\bibnamefont {Wright}}, \bibinfo {author} {\bibfnamefont
  {M.}~\bibnamefont {Zerkle}}, \bibinfo {author} {\bibfnamefont
  {G.}~\bibnamefont {Žerovnik}}, \ and\ \bibinfo {author} {\bibfnamefont
  {Y.}~\bibnamefont {Zhu}},\ }\href {\doibase 10.1016/j.nds.2018.02.001}
  {\bibfield  {journal} {\bibinfo  {journal} {Nuclear Data Sheets}\ }\textbf
  {\bibinfo {volume} {148}},\ \bibinfo {pages} {1} (\bibinfo {year}
  {2018})}\BibitemShut {NoStop}%
\bibitem [{\citenamefont {Kellett}\ \emph {et~al.}(2009)\citenamefont
  {Kellett}, \citenamefont {Bersillon},\ and\ \citenamefont
  {Mills}}]{kellett_jeff-31-311_2009}%
  \BibitemOpen
  \bibfield  {author} {\bibinfo {author} {\bibfnamefont {M.~A.}\ \bibnamefont
  {Kellett}}, \bibinfo {author} {\bibfnamefont {O.}~\bibnamefont {Bersillon}},
  \ and\ \bibinfo {author} {\bibfnamefont {R.}~\bibnamefont {Mills}},\
  }\href@noop {} {\  (\bibinfo {year} {2009})}\BibitemShut {NoStop}%
\bibitem [{\citenamefont {Shibata}\ \emph {et~al.}(2011)\citenamefont
  {Shibata}, \citenamefont {Iwamoto}, \citenamefont {Nakagawa}, \citenamefont
  {Iwamoto}, \citenamefont {Ichihara}, \citenamefont {Kunieda}, \citenamefont
  {Chiba}, \citenamefont {Furutaka}, \citenamefont {Otuka}, \citenamefont
  {Ohsawa}, \citenamefont {Murata}, \citenamefont {Matsunobu}, \citenamefont
  {Zukeran}, \citenamefont {Kamada},\ and\ \citenamefont
  {Katakura}}]{shibata_jendl-40_2011}%
  \BibitemOpen
  \bibfield  {author} {\bibinfo {author} {\bibfnamefont {K.}~\bibnamefont
  {Shibata}}, \bibinfo {author} {\bibfnamefont {O.}~\bibnamefont {Iwamoto}},
  \bibinfo {author} {\bibfnamefont {T.}~\bibnamefont {Nakagawa}}, \bibinfo
  {author} {\bibfnamefont {N.}~\bibnamefont {Iwamoto}}, \bibinfo {author}
  {\bibfnamefont {A.}~\bibnamefont {Ichihara}}, \bibinfo {author}
  {\bibfnamefont {S.}~\bibnamefont {Kunieda}}, \bibinfo {author} {\bibfnamefont
  {S.}~\bibnamefont {Chiba}}, \bibinfo {author} {\bibfnamefont
  {K.}~\bibnamefont {Furutaka}}, \bibinfo {author} {\bibfnamefont
  {N.}~\bibnamefont {Otuka}}, \bibinfo {author} {\bibfnamefont
  {T.}~\bibnamefont {Ohsawa}}, \bibinfo {author} {\bibfnamefont
  {T.}~\bibnamefont {Murata}}, \bibinfo {author} {\bibfnamefont
  {H.}~\bibnamefont {Matsunobu}}, \bibinfo {author} {\bibfnamefont
  {A.}~\bibnamefont {Zukeran}}, \bibinfo {author} {\bibfnamefont
  {S.}~\bibnamefont {Kamada}}, \ and\ \bibinfo {author} {\bibfnamefont {J.-i.}\
  \bibnamefont {Katakura}},\ }\href {\doibase 10.1080/18811248.2011.9711675}
  {\bibfield  {journal} {\bibinfo  {journal} {Journal of Nuclear Science and
  Technology}\ }\textbf {\bibinfo {volume} {48}},\ \bibinfo {pages} {1}
  (\bibinfo {year} {2011})}\BibitemShut {NoStop}%
\bibitem [{\citenamefont {Bulgac}\ \emph {et~al.}(2016)\citenamefont {Bulgac},
  \citenamefont {Magierski}, \citenamefont {Roche},\ and\ \citenamefont
  {Stetcu}}]{bulgac_induced_2016}%
  \BibitemOpen
  \bibfield  {author} {\bibinfo {author} {\bibfnamefont {A.}~\bibnamefont
  {Bulgac}}, \bibinfo {author} {\bibfnamefont {P.}~\bibnamefont {Magierski}},
  \bibinfo {author} {\bibfnamefont {K.~J.}\ \bibnamefont {Roche}}, \ and\
  \bibinfo {author} {\bibfnamefont {I.}~\bibnamefont {Stetcu}},\ }\href
  {\doibase 10.1103/PhysRevLett.116.122504} {\bibfield  {journal} {\bibinfo
  {journal} {Physical Review Letters}\ }\textbf {\bibinfo {volume} {116}},\
  \bibinfo {pages} {122504} (\bibinfo {year} {2016})}\BibitemShut {NoStop}%
\bibitem [{\citenamefont {Regnier}\ \emph {et~al.}(2016)\citenamefont
  {Regnier}, \citenamefont {Dubray}, \citenamefont {Schunck},\ and\
  \citenamefont {Verrière}}]{regnier_fission_2016}%
  \BibitemOpen
  \bibfield  {author} {\bibinfo {author} {\bibfnamefont {D.}~\bibnamefont
  {Regnier}}, \bibinfo {author} {\bibfnamefont {N.}~\bibnamefont {Dubray}},
  \bibinfo {author} {\bibfnamefont {N.}~\bibnamefont {Schunck}}, \ and\
  \bibinfo {author} {\bibfnamefont {M.}~\bibnamefont {Verrière}},\ }\href
  {\doibase 10.1103/PhysRevC.93.054611} {\bibfield  {journal} {\bibinfo
  {journal} {Physical Review C}\ }\textbf {\bibinfo {volume} {93}},\ \bibinfo
  {pages} {054611} (\bibinfo {year} {2016})},\ \bibinfo {note} {publisher:
  American Physical Society}\BibitemShut {NoStop}%
\bibitem [{\citenamefont {Younes}\ \emph {et~al.}(2019)\citenamefont {Younes},
  \citenamefont {Gogny},\ and\ \citenamefont
  {Berger}}]{younes_microscopic_2019}%
  \BibitemOpen
  \bibfield  {author} {\bibinfo {author} {\bibfnamefont {W.}~\bibnamefont
  {Younes}}, \bibinfo {author} {\bibfnamefont {D.~M.}\ \bibnamefont {Gogny}}, \
  and\ \bibinfo {author} {\bibfnamefont {J.-F.}\ \bibnamefont {Berger}},\
  }\href@noop {} {\emph {\bibinfo {title} {A microscopic theory of fission
  dynamics based on the generator coordinate method}}},\ Vol.\ \bibinfo
  {volume} {950}\ (\bibinfo  {publisher} {Springer},\ \bibinfo {year}
  {2019})\BibitemShut {NoStop}%
\bibitem [{\citenamefont {Randrup}\ and\ \citenamefont
  {Möller}(2011)}]{randrup_brownian_2011}%
  \BibitemOpen
  \bibfield  {author} {\bibinfo {author} {\bibfnamefont {J.}~\bibnamefont
  {Randrup}}\ and\ \bibinfo {author} {\bibfnamefont {P.}~\bibnamefont
  {Möller}},\ }\href {\doibase 10.1103/PhysRevLett.106.132503} {\bibfield
  {journal} {\bibinfo  {journal} {Physical Review Letters}\ }\textbf {\bibinfo
  {volume} {106}},\ \bibinfo {pages} {132503} (\bibinfo {year} {2011})},\
  \bibinfo {note} {publisher: American Physical Society}\BibitemShut {NoStop}%
\bibitem [{\citenamefont {Randrup}\ \emph {et~al.}(2011)\citenamefont
  {Randrup}, \citenamefont {Möller},\ and\ \citenamefont
  {Sierk}}]{randrup_fission-fragment_2011}%
  \BibitemOpen
  \bibfield  {author} {\bibinfo {author} {\bibfnamefont {J.}~\bibnamefont
  {Randrup}}, \bibinfo {author} {\bibfnamefont {P.}~\bibnamefont {Möller}}, \
  and\ \bibinfo {author} {\bibfnamefont {A.~J.}\ \bibnamefont {Sierk}},\ }\href
  {\doibase 10.1103/PhysRevC.84.034613} {\bibfield  {journal} {\bibinfo
  {journal} {Physical Review C}\ }\textbf {\bibinfo {volume} {84}},\ \bibinfo
  {pages} {034613} (\bibinfo {year} {2011})}\BibitemShut {NoStop}%
\bibitem [{\citenamefont {Pomorski}\ \emph {et~al.}(2017)\citenamefont
  {Pomorski}, \citenamefont {Ivanyuk},\ and\ \citenamefont
  {Nerlo-Pomorska}}]{pomorski_mass_2017}%
  \BibitemOpen
  \bibfield  {author} {\bibinfo {author} {\bibfnamefont {K.}~\bibnamefont
  {Pomorski}}, \bibinfo {author} {\bibfnamefont {F.~A.}\ \bibnamefont
  {Ivanyuk}}, \ and\ \bibinfo {author} {\bibfnamefont {B.}~\bibnamefont
  {Nerlo-Pomorska}},\ }\href {\doibase 10.1140/epja/i2017-12250-5} {\bibfield
  {journal} {\bibinfo  {journal} {The European Physical Journal A}\ }\textbf
  {\bibinfo {volume} {53}},\ \bibinfo {pages} {59} (\bibinfo {year}
  {2017})}\BibitemShut {NoStop}%
\bibitem [{\citenamefont {Liu}\ \emph {et~al.}(2019)\citenamefont {Liu},
  \citenamefont {Wu}, \citenamefont {Chen}, \citenamefont {Shen}, \citenamefont
  {Li},\ and\ \citenamefont {Ge}}]{liu_study_2019}%
  \BibitemOpen
  \bibfield  {author} {\bibinfo {author} {\bibfnamefont {L.-L.}\ \bibnamefont
  {Liu}}, \bibinfo {author} {\bibfnamefont {X.-Z.}\ \bibnamefont {Wu}},
  \bibinfo {author} {\bibfnamefont {Y.-J.}\ \bibnamefont {Chen}}, \bibinfo
  {author} {\bibfnamefont {C.-W.}\ \bibnamefont {Shen}}, \bibinfo {author}
  {\bibfnamefont {Z.-X.}\ \bibnamefont {Li}}, \ and\ \bibinfo {author}
  {\bibfnamefont {Z.-G.}\ \bibnamefont {Ge}},\ }\href {\doibase
  10.1103/PhysRevC.99.044614} {\bibfield  {journal} {\bibinfo  {journal}
  {Physical Review C}\ }\textbf {\bibinfo {volume} {99}},\ \bibinfo {pages}
  {044614} (\bibinfo {year} {2019})}\BibitemShut {NoStop}%
\bibitem [{\citenamefont {Fang}\ \emph {et~al.}(2021)\citenamefont {Fang},
  \citenamefont {Yu}, \citenamefont {Huang}, \citenamefont {Chen},
  \citenamefont {Su},\ and\ \citenamefont {Zhu}}]{fang_theoretical_2021}%
  \BibitemOpen
  \bibfield  {author} {\bibinfo {author} {\bibfnamefont {Z.-X.}\ \bibnamefont
  {Fang}}, \bibinfo {author} {\bibfnamefont {M.}~\bibnamefont {Yu}}, \bibinfo
  {author} {\bibfnamefont {Y.-G.}\ \bibnamefont {Huang}}, \bibinfo {author}
  {\bibfnamefont {J.-B.}\ \bibnamefont {Chen}}, \bibinfo {author}
  {\bibfnamefont {J.}~\bibnamefont {Su}}, \ and\ \bibinfo {author}
  {\bibfnamefont {L.}~\bibnamefont {Zhu}},\ }\href {\doibase
  10.1007/s41365-021-00911-0} {\bibfield  {journal} {\bibinfo  {journal}
  {Nuclear Science and Techniques}\ }\textbf {\bibinfo {volume} {32}},\
  \bibinfo {pages} {72} (\bibinfo {year} {2021})}\BibitemShut {NoStop}%
\bibitem [{\citenamefont {Brosa}\ \emph {et~al.}(1990)\citenamefont {Brosa},
  \citenamefont {Grossmann},\ and\ \citenamefont
  {Müller}}]{brosa_nuclear_1990}%
  \BibitemOpen
  \bibfield  {author} {\bibinfo {author} {\bibfnamefont {U.}~\bibnamefont
  {Brosa}}, \bibinfo {author} {\bibfnamefont {S.}~\bibnamefont {Grossmann}}, \
  and\ \bibinfo {author} {\bibfnamefont {A.}~\bibnamefont {Müller}},\ }\href
  {\doibase 10.1016/0370-1573(90)90114-H} {\bibfield  {journal} {\bibinfo
  {journal} {Physics Reports}\ }\textbf {\bibinfo {volume} {197}},\ \bibinfo
  {pages} {167} (\bibinfo {year} {1990})}\BibitemShut {NoStop}%
\bibitem [{\citenamefont {Schmidt}\ \emph {et~al.}(2016)\citenamefont
  {Schmidt}, \citenamefont {Jurado}, \citenamefont {Amouroux},\ and\
  \citenamefont {Schmitt}}]{schmidt_general_2016}%
  \BibitemOpen
  \bibfield  {author} {\bibinfo {author} {\bibfnamefont {K.-H.}\ \bibnamefont
  {Schmidt}}, \bibinfo {author} {\bibfnamefont {B.}~\bibnamefont {Jurado}},
  \bibinfo {author} {\bibfnamefont {C.}~\bibnamefont {Amouroux}}, \ and\
  \bibinfo {author} {\bibfnamefont {C.}~\bibnamefont {Schmitt}},\ }\href
  {\doibase 10.1016/j.nds.2015.12.009} {\bibfield  {journal} {\bibinfo
  {journal} {Nuclear Data Sheets}\ }\textbf {\bibinfo {volume} {131}},\
  \bibinfo {pages} {107} (\bibinfo {year} {2016})}\BibitemShut {NoStop}%
\bibitem [{\citenamefont {Schmidt}\ and\ \citenamefont
  {Jurado}(2018)}]{schmidt2018review}%
  \BibitemOpen
  \bibfield  {author} {\bibinfo {author} {\bibfnamefont {K.-H.}\ \bibnamefont
  {Schmidt}}\ and\ \bibinfo {author} {\bibfnamefont {B.}~\bibnamefont
  {Jurado}},\ }\href@noop {} {\bibfield  {journal} {\bibinfo  {journal}
  {Reports on Progress in Physics}\ }\textbf {\bibinfo {volume} {81}},\
  \bibinfo {pages} {106301} (\bibinfo {year} {2018})}\BibitemShut {NoStop}%
\bibitem [{\citenamefont {Wang}\ \emph {et~al.}(2019)\citenamefont {Wang},
  \citenamefont {Pei}, \citenamefont {Liu},\ and\ \citenamefont
  {Qiang}}]{wang_bayesian_2019}%
  \BibitemOpen
  \bibfield  {author} {\bibinfo {author} {\bibfnamefont {Z.-A.}\ \bibnamefont
  {Wang}}, \bibinfo {author} {\bibfnamefont {J.}~\bibnamefont {Pei}}, \bibinfo
  {author} {\bibfnamefont {Y.}~\bibnamefont {Liu}}, \ and\ \bibinfo {author}
  {\bibfnamefont {Y.}~\bibnamefont {Qiang}},\ }\href {\doibase
  10.1103/PhysRevLett.123.122501} {\bibfield  {journal} {\bibinfo  {journal}
  {Physical Review Letters}\ }\textbf {\bibinfo {volume} {123}},\ \bibinfo
  {pages} {122501} (\bibinfo {year} {2019})}\BibitemShut {NoStop}%
\bibitem [{\citenamefont {Niu}\ and\ \citenamefont
  {Liang}(2018)}]{niu_nuclear_2018}%
  \BibitemOpen
  \bibfield  {author} {\bibinfo {author} {\bibfnamefont {Z.}~\bibnamefont
  {Niu}}\ and\ \bibinfo {author} {\bibfnamefont {H.}~\bibnamefont {Liang}},\
  }\href {\doibase 10.1016/j.physletb.2018.01.002} {\bibfield  {journal}
  {\bibinfo  {journal} {Physics Letters B}\ }\textbf {\bibinfo {volume}
  {778}},\ \bibinfo {pages} {48} (\bibinfo {year} {2018})}\BibitemShut
  {NoStop}%
\bibitem [{\citenamefont {Utama}\ \emph
  {et~al.}(2016{\natexlab{a}})\citenamefont {Utama}, \citenamefont
  {Piekarewicz},\ and\ \citenamefont {Prosper}}]{utama_nuclear_2016-1}%
  \BibitemOpen
  \bibfield  {author} {\bibinfo {author} {\bibfnamefont {R.}~\bibnamefont
  {Utama}}, \bibinfo {author} {\bibfnamefont {J.}~\bibnamefont {Piekarewicz}},
  \ and\ \bibinfo {author} {\bibfnamefont {H.~B.}\ \bibnamefont {Prosper}},\
  }\href {\doibase 10.1103/PhysRevC.93.014311} {\bibfield  {journal} {\bibinfo
  {journal} {Physical Review C}\ }\textbf {\bibinfo {volume} {93}},\ \bibinfo
  {pages} {014311} (\bibinfo {year} {2016}{\natexlab{a}})}\BibitemShut
  {NoStop}%
\bibitem [{\citenamefont {Utama}\ \emph
  {et~al.}(2016{\natexlab{b}})\citenamefont {Utama}, \citenamefont {Chen},\
  and\ \citenamefont {Piekarewicz}}]{utama_nuclear_2016}%
  \BibitemOpen
  \bibfield  {author} {\bibinfo {author} {\bibfnamefont {R.}~\bibnamefont
  {Utama}}, \bibinfo {author} {\bibfnamefont {W.-C.}\ \bibnamefont {Chen}}, \
  and\ \bibinfo {author} {\bibfnamefont {J.}~\bibnamefont {Piekarewicz}},\
  }\href {\doibase 10.1088/0954-3899/43/11/114002} {\bibfield  {journal}
  {\bibinfo  {journal} {Journal of Physics G: Nuclear and Particle Physics}\
  }\textbf {\bibinfo {volume} {43}},\ \bibinfo {pages} {114002} (\bibinfo
  {year} {2016}{\natexlab{b}})}\BibitemShut {NoStop}%
\bibitem [{\citenamefont {Ma}\ \emph {et~al.}(2020)\citenamefont {Ma},
  \citenamefont {Peng}, \citenamefont {Wei}, \citenamefont {Niu}, \citenamefont
  {Wang},\ and\ \citenamefont {Wada}}]{ma_isotopic_2020}%
  \BibitemOpen
  \bibfield  {author} {\bibinfo {author} {\bibfnamefont {C.-W.}\ \bibnamefont
  {Ma}}, \bibinfo {author} {\bibfnamefont {D.}~\bibnamefont {Peng}}, \bibinfo
  {author} {\bibfnamefont {H.-L.}\ \bibnamefont {Wei}}, \bibinfo {author}
  {\bibfnamefont {Z.-M.}\ \bibnamefont {Niu}}, \bibinfo {author} {\bibfnamefont
  {Y.-T.}\ \bibnamefont {Wang}}, \ and\ \bibinfo {author} {\bibfnamefont
  {R.}~\bibnamefont {Wada}},\ }\href {\doibase 10.1088/1674-1137/44/1/014104}
  {\bibfield  {journal} {\bibinfo  {journal} {Chinese Physics C}\ }\textbf
  {\bibinfo {volume} {44}},\ \bibinfo {pages} {014104} (\bibinfo {year}
  {2020})}\BibitemShut {NoStop}%
\bibitem [{\citenamefont {Song}\ \emph {et~al.}(2022)\citenamefont {Song},
  \citenamefont {Zhu},\ and\ \citenamefont {Su}}]{song2022target}%
  \BibitemOpen
  \bibfield  {author} {\bibinfo {author} {\bibfnamefont {Q.-F.}\ \bibnamefont
  {Song}}, \bibinfo {author} {\bibfnamefont {L.}~\bibnamefont {Zhu}}, \ and\
  \bibinfo {author} {\bibfnamefont {J.}~\bibnamefont {Su}},\ }\href {\doibase
  10.1088/1674-1137/ac6249} {\bibfield  {journal} {\bibinfo  {journal} {Chinese
  Physics C}\ }\textbf {\bibinfo {volume} {46}},\ \bibinfo {pages} {074108}
  (\bibinfo {year} {2022})}\BibitemShut {NoStop}%
\bibitem [{\citenamefont {Neufcourt}\ \emph {et~al.}(2019)\citenamefont
  {Neufcourt}, \citenamefont {Cao}, \citenamefont {Nazarewicz}, \citenamefont
  {Olsen},\ and\ \citenamefont {Viens}}]{neufcourt_neutron_2019}%
  \BibitemOpen
  \bibfield  {author} {\bibinfo {author} {\bibfnamefont {L.}~\bibnamefont
  {Neufcourt}}, \bibinfo {author} {\bibfnamefont {Y.}~\bibnamefont {Cao}},
  \bibinfo {author} {\bibfnamefont {W.}~\bibnamefont {Nazarewicz}}, \bibinfo
  {author} {\bibfnamefont {E.}~\bibnamefont {Olsen}}, \ and\ \bibinfo {author}
  {\bibfnamefont {F.}~\bibnamefont {Viens}},\ }\href {\doibase
  10.1103/PhysRevLett.122.062502} {\bibfield  {journal} {\bibinfo  {journal}
  {Physical Review Letters}\ }\textbf {\bibinfo {volume} {122}},\ \bibinfo
  {pages} {062502} (\bibinfo {year} {2019})}\BibitemShut {NoStop}%
\bibitem [{\citenamefont {Neufcourt}\ \emph {et~al.}(2020)\citenamefont
  {Neufcourt}, \citenamefont {Cao}, \citenamefont {Giuliani}, \citenamefont
  {Nazarewicz}, \citenamefont {Olsen},\ and\ \citenamefont
  {Tarasov}}]{neufcourt_beyond_2020}%
  \BibitemOpen
  \bibfield  {author} {\bibinfo {author} {\bibfnamefont {L.}~\bibnamefont
  {Neufcourt}}, \bibinfo {author} {\bibfnamefont {Y.}~\bibnamefont {Cao}},
  \bibinfo {author} {\bibfnamefont {S.}~\bibnamefont {Giuliani}}, \bibinfo
  {author} {\bibfnamefont {W.}~\bibnamefont {Nazarewicz}}, \bibinfo {author}
  {\bibfnamefont {E.}~\bibnamefont {Olsen}}, \ and\ \bibinfo {author}
  {\bibfnamefont {O.~B.}\ \bibnamefont {Tarasov}},\ }\href {\doibase
  10.1103/PhysRevC.101.014319} {\bibfield  {journal} {\bibinfo  {journal}
  {Physical Review C}\ }\textbf {\bibinfo {volume} {101}},\ \bibinfo {pages}
  {014319} (\bibinfo {year} {2020})}\BibitemShut {NoStop}%
\bibitem [{\citenamefont {Lovell}\ \emph {et~al.}(2019)\citenamefont {Lovell},
  \citenamefont {Mohan}, \citenamefont {Talou},\ and\ \citenamefont
  {Chertkov}}]{lovell_constraining_2019}%
  \BibitemOpen
  \bibfield  {author} {\bibinfo {author} {\bibfnamefont {A.}~\bibnamefont
  {Lovell}}, \bibinfo {author} {\bibfnamefont {A.}~\bibnamefont {Mohan}},
  \bibinfo {author} {\bibfnamefont {P.}~\bibnamefont {Talou}}, \ and\ \bibinfo
  {author} {\bibfnamefont {M.}~\bibnamefont {Chertkov}},\ }\href {\doibase
  10.1051/epjconf/201921104006} {\bibfield  {journal} {\bibinfo  {journal} {EPJ
  Web of Conferences}\ }\textbf {\bibinfo {volume} {211}},\ \bibinfo {pages}
  {04006} (\bibinfo {year} {2019})}\BibitemShut {NoStop}%
\bibitem [{\citenamefont {Qiao}\ \emph {et~al.}(2021)\citenamefont {Qiao},
  \citenamefont {Pei}, \citenamefont {Wang}, \citenamefont {Qiang},
  \citenamefont {Chen}, \citenamefont {Shu},\ and\ \citenamefont
  {Ge}}]{qiao_bayesian_2021}%
  \BibitemOpen
  \bibfield  {author} {\bibinfo {author} {\bibfnamefont {C.~Y.}\ \bibnamefont
  {Qiao}}, \bibinfo {author} {\bibfnamefont {J.~C.}\ \bibnamefont {Pei}},
  \bibinfo {author} {\bibfnamefont {Z.~A.}\ \bibnamefont {Wang}}, \bibinfo
  {author} {\bibfnamefont {Y.}~\bibnamefont {Qiang}}, \bibinfo {author}
  {\bibfnamefont {Y.~J.}\ \bibnamefont {Chen}}, \bibinfo {author}
  {\bibfnamefont {N.~C.}\ \bibnamefont {Shu}}, \ and\ \bibinfo {author}
  {\bibfnamefont {Z.~G.}\ \bibnamefont {Ge}},\ }\href {\doibase
  10.1103/PhysRevC.103.034621} {\bibfield  {journal} {\bibinfo  {journal}
  {Physical Review C}\ }\textbf {\bibinfo {volume} {103}},\ \bibinfo {pages}
  {034621} (\bibinfo {year} {2021})}\BibitemShut {NoStop}%
\bibitem [{\citenamefont {Liu}\ \emph {et~al.}(2013)\citenamefont {Liu},
  \citenamefont {Musialski}, \citenamefont {Wonka},\ and\ \citenamefont
  {Ye}}]{liu_tensor_2013}%
  \BibitemOpen
  \bibfield  {author} {\bibinfo {author} {\bibfnamefont {J.}~\bibnamefont
  {Liu}}, \bibinfo {author} {\bibfnamefont {P.}~\bibnamefont {Musialski}},
  \bibinfo {author} {\bibfnamefont {P.}~\bibnamefont {Wonka}}, \ and\ \bibinfo
  {author} {\bibfnamefont {J.}~\bibnamefont {Ye}},\ }\href {\doibase
  10.1109/TPAMI.2012.39} {\bibfield  {journal} {\bibinfo  {journal} {IEEE
  Transactions on Pattern Analysis and Machine Intelligence}\ }\textbf
  {\bibinfo {volume} {35}},\ \bibinfo {pages} {208} (\bibinfo {year}
  {2013})}\BibitemShut {NoStop}%
\bibitem [{\citenamefont {Chen}\ \emph {et~al.}(2019)\citenamefont {Chen},
  \citenamefont {He},\ and\ \citenamefont {Sun}}]{chen_bayesian_2019}%
  \BibitemOpen
  \bibfield  {author} {\bibinfo {author} {\bibfnamefont {X.}~\bibnamefont
  {Chen}}, \bibinfo {author} {\bibfnamefont {Z.}~\bibnamefont {He}}, \ and\
  \bibinfo {author} {\bibfnamefont {L.}~\bibnamefont {Sun}},\ }\href {\doibase
  10.1016/j.trc.2018.11.003} {\bibfield  {journal} {\bibinfo  {journal}
  {Transportation Research Part C: Emerging Technologies}\ }\textbf {\bibinfo
  {volume} {98}},\ \bibinfo {pages} {73} (\bibinfo {year} {2019})}\BibitemShut
  {NoStop}%
\bibitem [{\citenamefont {Chen}\ \emph {et~al.}(2017)\citenamefont {Chen},
  \citenamefont {Han}, \citenamefont {Wang}, \citenamefont {Zhao},
  \citenamefont {Meng}, \citenamefont {Lin},\ and\ \citenamefont
  {Tang}}]{chen_general_2015}%
  \BibitemOpen
  \bibfield  {author} {\bibinfo {author} {\bibfnamefont {X.}~\bibnamefont
  {Chen}}, \bibinfo {author} {\bibfnamefont {Z.}~\bibnamefont {Han}}, \bibinfo
  {author} {\bibfnamefont {Y.}~\bibnamefont {Wang}}, \bibinfo {author}
  {\bibfnamefont {Q.}~\bibnamefont {Zhao}}, \bibinfo {author} {\bibfnamefont
  {D.}~\bibnamefont {Meng}}, \bibinfo {author} {\bibfnamefont {L.}~\bibnamefont
  {Lin}}, \ and\ \bibinfo {author} {\bibfnamefont {Y.}~\bibnamefont {Tang}},\
  }\href@noop {} {\bibfield  {journal} {\bibinfo  {journal} {arXiv preprint
  arXiv:1705.06755}\ } (\bibinfo {year} {2017})}\BibitemShut {NoStop}%
\bibitem [{\citenamefont {Koning}\ and\ \citenamefont
  {Rochman}(2012)}]{koning_modern_2012}%
  \BibitemOpen
  \bibfield  {author} {\bibinfo {author} {\bibfnamefont {A.~J.}\ \bibnamefont
  {Koning}}\ and\ \bibinfo {author} {\bibfnamefont {D.}~\bibnamefont
  {Rochman}},\ }\href {\doibase 10.1016/j.nds.2012.11.002} {\bibfield
  {journal} {\bibinfo  {journal} {Nuclear Data Sheets}\ }\bibinfo {series}
  {Special {Issue} on {Nuclear} {Reaction} {Data}},\ \textbf {\bibinfo {volume}
  {113}},\ \bibinfo {pages} {2841} (\bibinfo {year} {2012})}\BibitemShut
  {NoStop}%
\bibitem [{\citenamefont {Lindner}\ and\ \citenamefont
  {Seegmiller}(1990)}]{lindner_reactor_1990}%
  \BibitemOpen
  \bibfield  {author} {\bibinfo {author} {\bibfnamefont {M.}~\bibnamefont
  {Lindner}}\ and\ \bibinfo {author} {\bibfnamefont {D.~W.}\ \bibnamefont
  {Seegmiller}},\ }\href@noop {} {\bibfield  {journal} {\bibinfo  {journal}
  {Radiochimica Acta}\ }\textbf {\bibinfo {volume} {49}},\ \bibinfo {pages} {1}
  (\bibinfo {year} {1990})}\BibitemShut {NoStop}%
\bibitem [{\citenamefont {Gindler}\ \emph {et~al.}(1981)\citenamefont
  {Gindler}, \citenamefont {Glendenin}, \citenamefont {Krapp}, \citenamefont
  {Fernandez}, \citenamefont {Flynn},\ and\ \citenamefont
  {Henderson}}]{gindler_mass_nodate}%
  \BibitemOpen
  \bibfield  {author} {\bibinfo {author} {\bibfnamefont {J.}~\bibnamefont
  {Gindler}}, \bibinfo {author} {\bibfnamefont {L.}~\bibnamefont {Glendenin}},
  \bibinfo {author} {\bibfnamefont {E.}~\bibnamefont {Krapp}}, \bibinfo
  {author} {\bibfnamefont {S.}~\bibnamefont {Fernandez}}, \bibinfo {author}
  {\bibfnamefont {K.}~\bibnamefont {Flynn}}, \ and\ \bibinfo {author}
  {\bibfnamefont {D.}~\bibnamefont {Henderson}},\ }\href@noop {} {\bibfield
  {journal} {\bibinfo  {journal} {Journal of Inorganic and Nuclear Chemistry}\
  }\textbf {\bibinfo {volume} {43}},\ \bibinfo {pages} {445} (\bibinfo {year}
  {1981})}\BibitemShut {NoStop}%
\bibitem [{\citenamefont {Glendenin}\ \emph {et~al.}(1981)\citenamefont
  {Glendenin}, \citenamefont {Gindler}, \citenamefont {Henderson},\ and\
  \citenamefont {Meadows}}]{glendenin_mass_1981}%
  \BibitemOpen
  \bibfield  {author} {\bibinfo {author} {\bibfnamefont {L.~E.}\ \bibnamefont
  {Glendenin}}, \bibinfo {author} {\bibfnamefont {J.~E.}\ \bibnamefont
  {Gindler}}, \bibinfo {author} {\bibfnamefont {D.~J.}\ \bibnamefont
  {Henderson}}, \ and\ \bibinfo {author} {\bibfnamefont {J.~W.}\ \bibnamefont
  {Meadows}},\ }\href {\doibase 10.1103/PhysRevC.24.2600} {\bibfield  {journal}
  {\bibinfo  {journal} {Physical Review C}\ }\textbf {\bibinfo {volume} {24}},\
  \bibinfo {pages} {2600} (\bibinfo {year} {1981})}\BibitemShut {NoStop}%
\bibitem [{\citenamefont {Agarwal}\ \emph {et~al.}(2008)\citenamefont
  {Agarwal}, \citenamefont {Goswami}, \citenamefont {Kalsi}, \citenamefont
  {Singh}, \citenamefont {Mhatre},\ and\ \citenamefont
  {Ramaswami}}]{agarwal_mass_2008}%
  \BibitemOpen
  \bibfield  {author} {\bibinfo {author} {\bibfnamefont {C.}~\bibnamefont
  {Agarwal}}, \bibinfo {author} {\bibfnamefont {A.}~\bibnamefont {Goswami}},
  \bibinfo {author} {\bibfnamefont {P.~C.}\ \bibnamefont {Kalsi}}, \bibinfo
  {author} {\bibfnamefont {S.}~\bibnamefont {Singh}}, \bibinfo {author}
  {\bibfnamefont {A.}~\bibnamefont {Mhatre}}, \ and\ \bibinfo {author}
  {\bibfnamefont {A.}~\bibnamefont {Ramaswami}},\ }\href {\doibase
  10.1007/s10967-007-7011-8} {\bibfield  {journal} {\bibinfo  {journal}
  {Journal of Radioanalytical and Nuclear Chemistry}\ }\textbf {\bibinfo
  {volume} {275}},\ \bibinfo {pages} {445} (\bibinfo {year}
  {2008})}\BibitemShut {NoStop}%
\end{thebibliography}%

\end{document}